\documentclass[%
                        reprint,
                        nofootinbib,
                        amsmath,amssymb,
                        aps,
                        floatfix,
                        ]{revtex4-2}
                        \usepackage{cleveref}
                        \usepackage{graphicx}
                        \usepackage{dcolumn}
                        \usepackage{bm}
                        
         \usepackage{verbatim}
        
                       \usepackage{lipsum}%
                        \usepackage{wrapfig}
                        \usepackage{mathrsfs}  
                        \usepackage{feynmp-auto}

\begin{document}
                        
                        \preprint{APS/123-QED}
                        
                        \title{Single Pion Production off Free Nucleons: Analysis of Photon, Electron, Pion and Neutrino Induced Processes}
                        
    \author{M. Kabirnezhad}
    \affiliation{%
    Imperial College London, Department of Physics,London SW7 2BZ, United Kingdom\\
    York University, Department of Physics and Astronomy, 4700 Keele Street, Toronto M3J 1P3, Canada\\
    University of Oxford, Department of Physics, Oxford OX1 3RH, United Kingdom
    }
    \email{minoo.kabirnezhad@physics.ox.ac.uk}
                        
                        \date{\today}
                        
        \begin{abstract}
        I present a unified model for single pion production in photo-, electro-, and neutrino--nucleon interactions that is applicable across the broad kinematic range in the GeV regime relevant to accelerator-based neutrino experiments. The model incorporates vector and axial-vector transition form factors for excited nucleon states with masses up to $2~\text{GeV}$, together with non-resonant background contributions, within a meson-dominance framework that respects quantum chromodynamic constraints and preserves unitarity. This construction ensures the correct asymptotic behaviour at large momentum transfer ($Q^2$) while providing a consistent description of the resonance and transition regions. At very low $Q^2$, the implementation of the Conserved Vector Current and Partially Conserved Axial Current relations leads to reliable predictions and helps address challenges encountered in current neutrino data analyses.
        
        The unified framework enables a global analysis that combines the available electron-, photon-, pion-, and neutrino-scattering data. This comprehensive approach allows detailed studies of nucleon structure in the resonance region and provides important constraints on weak interaction form factors, where modern neutrino--nucleon data remain limited. By determining all model parameters simultaneously with their correlated uncertainties, the approach establishes a robust, data-driven foundation for precision neutrino interaction modelling in support of next-generation neutrino oscillation measurements.
        
        \end{abstract}
        
                \maketitle
                        
        \section{Introduction}
        
        Precise modelling of neutrino interactions \cite{Ankowski:2014yfa,Megias:2016lke,gonzalez-jimenez2019,Gonzalez-Jimenez:2019ejf,Sobczyk:2021dwm, nikolakopoulos2023, Lovato:2023raf,rocco2019} is essential for predicting both overall interaction rates and detailed final-state kinematics in neutrino experiments \cite{Nikolakopoulos:2022tut, McKean:2025khb, Dolan:2026dfu,Dolan:2026nlr}. These models determine how the true neutrino energy maps onto the observable energy deposited in a detector, a relationship that directly affects oscillation analyses and the control of systematic uncertainties. Achieving accurate cross-section predictions across the wide kinematic range probed by broad-band neutrino beams is therefore indispensable for meeting the precision goals of current and next-generation experiments.
        
        Among the processes relevant to this energy regime, neutrino interactions that produce a single pion in the final state play a central role. In the $1-3~\text{GeV}$ energy region, characteristic of all modern accelerator-based neutrino beams, single pion production (SPP) constitutes the largest component of the inclusive neutrino–nucleus cross section. A precise and theoretically robust description of SPP dynamics is therefore essential for reducing interaction-model systematics in oscillation measurements.
        
        Modelling SPP is inherently challenging because it involves two distinct but experimentally indistinguishable production mechanisms. The first is resonance production, in which the exchanged boson excites the nucleon to a baryon resonance that subsequently decays into a pion-nucleon final state. The second is non-resonant production, where the pion is created directly at the interaction vertex alongside the outgoing nucleon. These mechanisms produce final states that are kinematically identical and give rise to non-negligible interference effects. A reliable theoretical framework must therefore treat resonant and non-resonant amplitudes coherently, with a consistent phase structure across the entire kinematic domain.   
        
\begin{figure*}[t]
     \begin{center}	 \includegraphics[width=0.49\textwidth] {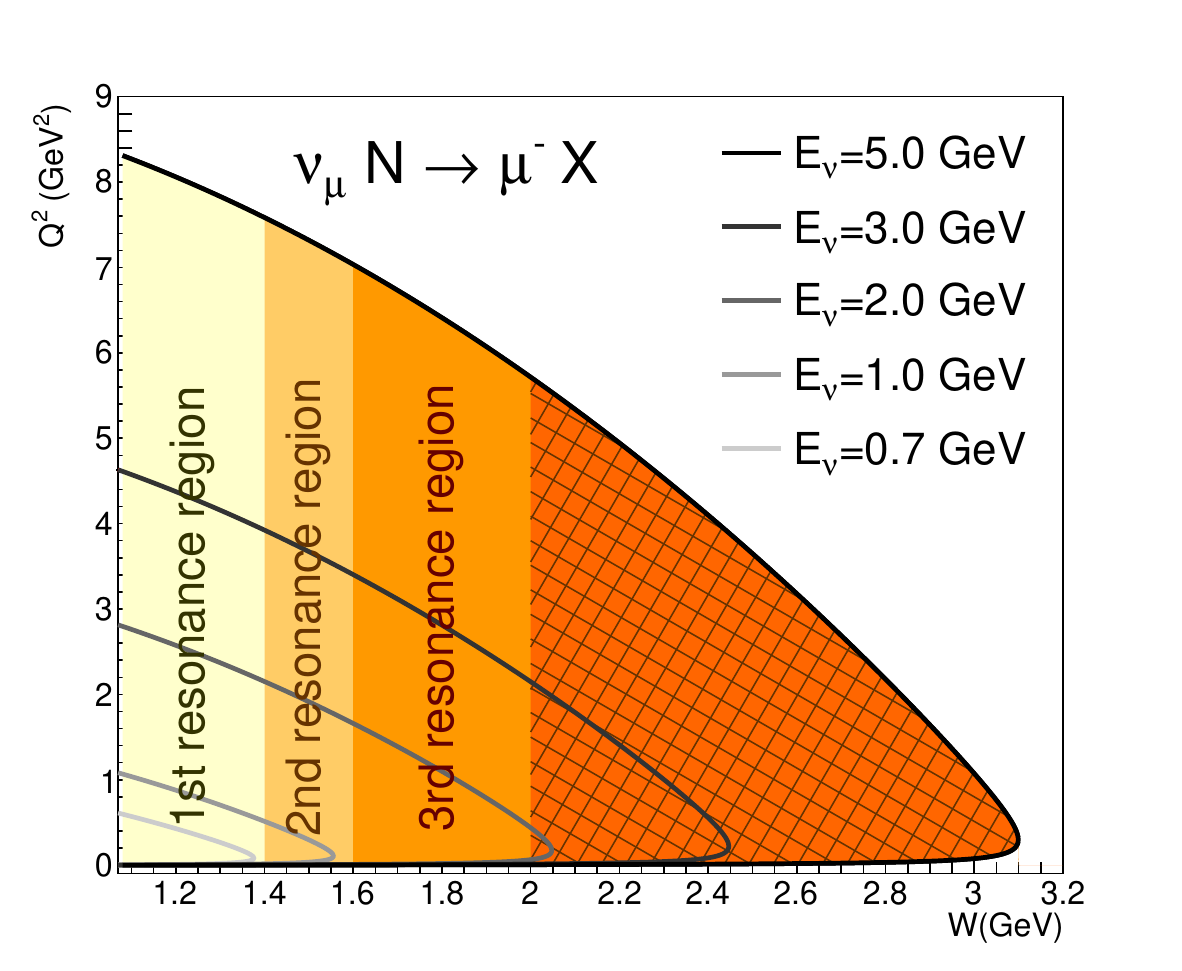}
    \includegraphics[width=0.49\textwidth] {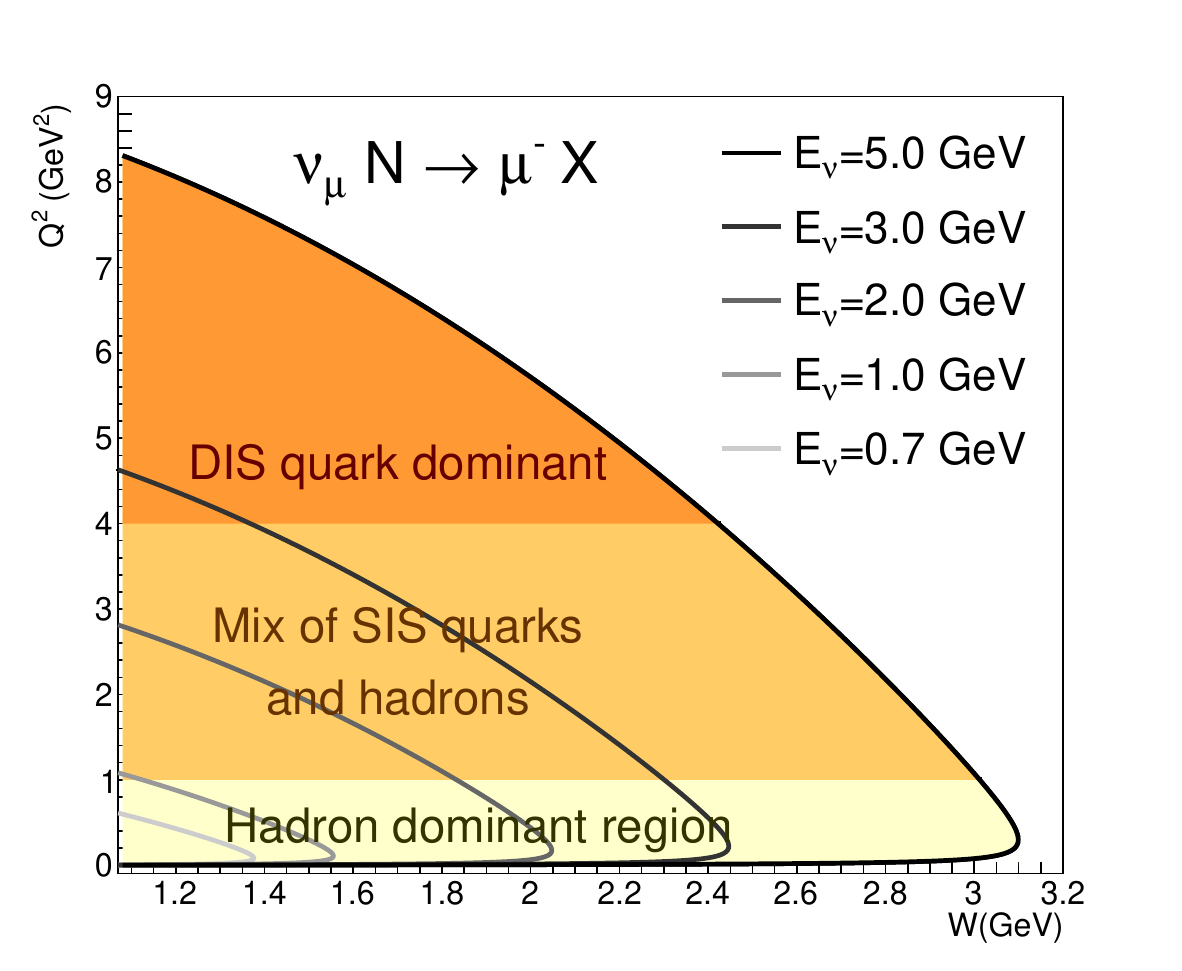}
        \caption{The allowed kinematic region for charged-current inelastic neutrino--nucleon interactions in terms of $Q^2$ and $W$ for different neutrino energies. The left panel illustrates the evolution in $W$ and highlights the different resonance regions. While the resonance region extends up to $W = 2.0~\text{GeV}$, single-pion production can still occur at higher $W$ through non-resonant interactions. The right panel shows the corresponding evolution in $Q^2$, following Ref.~\cite{Muzakka:2022wey}: at low $Q^2$ neutrinos predominantly probe the nucleon as a whole, whereas at higher $Q^2$ they resolve individual quarks, with a broad transition region between hadronic and partonic degrees of freedom.}   
                    \label{PhaseSpace2}
                \end{center}
                \end{figure*}

        Inelastic neutrino-nucleon scattering, including SPP channels, can be described using two independent kinematic variables, typically chosen as invariants: the four-momentum transfer ($Q^2$) and the invariant hadronic mass ($W$).
         The accessible kinematic region in terms of $Q^2$ and $W$
         for various neutrino energies relevant to current and future accelerator-based experiments is shown in Fig.~\ref{PhaseSpace2}. The neutrino energy spectra for T2K, Hyper-Kamiokande, and MicroBooNE peak below $E_{\nu}=1$ GeV, while NOvA operates primarily in the range of $1<E_{\nu}<3$ GeV, and DUNE extends up to $E_{\nu}<5~$GeV. Consequently, for wide-band beams in accelerator-based neutrino experiments, interaction models must remain valid across a broad range of $Q^2$ and $W$.
         
At low invariant mass $W$ ($<1.4$~GeV), the target nucleon is excited to the $\Delta(1232)$ resonance, which promptly decays into a nucleon and a pion. The $\Delta$ resonance provides the dominant contribution in the resonance region and is the sole resonance present in the first resonance region ($1.08~\text{GeV}<W<1.4$~GeV). At higher $W$, the second resonance region ($1.4~\text{GeV}<W<1.6$~GeV) contains three isospin-$1/2$ resonances: the $P_{11}(1440)$ Roper resonance and the two negative-parity states $D_{13}(1520)$ and $S_{11}(1535)$. Beyond this, the third resonance region ($1.6~\text{GeV}<W<2.0$~GeV) includes several additional resonances, although their overall contribution to the cross section is smaller. In both the second and third resonance regions, mesons heavier than the pion, such as $\eta$, $K$, $\omega$, and $\rho$, can also be produced. 
        
        The $Q^2$ dependence further characterises the scattering regime as illustrated in the right plot of Fig. \ref{PhaseSpace2}. At low $Q^2$ ($\lesssim 1$~GeV$^2$), the interaction is dominated by hadronic resonance production and non-resonant background processes, without quark-level interactions inside the nucleon. In the intermediate region ($1 \lesssim Q^2 \lesssim 4$~GeV$^2$), hadronic resonance production coexists with non-perturbative shallow inelastic scattering (SIS), often described as the quark–hadron transition region. At higher momentum transfers ($Q^2 \gtrsim 4$~GeV$^2$), perturbative quantum chromodynamic (QCD) becomes applicable, and deep inelastic scattering (DIS) processes, dominated by quark-level interactions, prevail. 
                
    In addition to model development and robust predictions, a crucial aspect of theoretical work in neutrino experiments is the quantification of uncertainties. While several SPP models exist \cite{Adler,RS,Rein,HNV,Mariano2011,Nakamura:2015rta,Alam,Gonzalez-Jimenez:2016qqq,Yao:2018pzc,Yao:2019avf,sobczyk:2014xza,proc2,proc3}, most of them address only key processes, focusing primarily on the first and second resonance regions. However, none of these models systematically study or evaluate the systematic uncertainties inherent to the models themselves. Furthermore, existing SPP models are confined to the non-perturbative region, which does not encompass the broad kinematic range required for neutrino experiments. The primary goal of this work is to develop a comprehensive model that addresses these gaps, providing a robust framework to enhance the precision of neutrino measurements.
    
    A commonly used parametrisation of nucleon form factors in theoretical models is the dipole form factor, originally derived from elastic electron-scattering data. However, measurements from inelastic scattering indicate that the dipole form fails to accurately describe excited nucleon states, particularly in the transition region. In contrast, the meson dominance (MD) form factor model \cite{Sakurai:1960ju, Gell-Mann:1961jim}, which accounts for the interaction between leptons and nucleons via meson exchange with analogous quantum properties,, provides a more realistic description~\cite{Lichard}. As emphasised in Ref.~\cite{stoler}, the MD framework, consistent with QCD expectations, successfully captures the behaviour of nucleon form factors across both perturbative and non-perturbative regimes, including the hadron-quark transition region. The model contains a set of parameters constrained by unitarity and analyticity considerations; however, their precise determination requires high-quality data spanning a wide kinematic range, which is presently limited by the scarcity of neutrino interaction measurements.
    
    To address this limitation, this work introduces a unified framework that extends the description of single pion production to analogous processes induced not only by neutrino beams but also by electron, photon, and pion probes. The form factors governing these interactions are fundamentally related. In particular, the vector-current form factors entering weak interactions are connected to proton and neutron electromagnetic form factors through isospin symmetry, as summarised in Table~\ref{Isospin_symmetry}. Consequently, combining information from different probes is especially advantageous, since electron-, photon-, and pion-scattering experiments provide extensive and precise datasets covering a broad kinematic domain. Exploiting these complementary measurements enables a more comprehensive investigation of nucleon structure than can be achieved using neutrino data alone.
    
    The unified model leverages the strengths of each interaction channel. Electron scattering data provide precise constraints on the vector current over a wide range of $Q^2$, while photon-induced reactions offer complementary sensitivity in the very low-$Q^2$ region. Pion scattering data, interpreted within the framework of the Partially Conserved Axial Current (PCAC) hypothesis, supply essential information on the axial-vector current at low $Q^2$. The combined dataset therefore provides a powerful foundation for addressing longstanding tensions between single-pion production models and neutrino measurements, particularly in the low-$Q^2$ region.
    
    \begin{table*}[t]
        \centering
            \caption{Isospin relation for vector form factors where $s_W=\left( \frac{1}{2}-2\sin^2\theta_W \right)$. Note that in the case of an isospin 1/2 → 3/2 transition, the form factors are equal for proton and neutron which is indicated by the index N (instead of p or n).  }
             \label{Isospin_symmetry}
        \begin{ruledtabular}
            \begin{tabular}{|l|l|l|l|l|}
            Channels~& $\mathcal{F}^V_i \text{ for } I=1/2$ & $\mathcal{C}^V_i \text{ for }I=1/2$&$\mathcal{F}^V_i \text{ for }I=3/2$&$\mathcal{C}^V_i \text{ for }I=3/2$\\[3pt]
                         \hline
            $\begin{aligned}
                &e^{-}p \rightarrow e^{-}R^{+}~\\[3pt]
                &e^{-}p \rightarrow e^{-}R^{0}~\\[3pt]
                &\nu p\rightarrow l^{-}R^{++}~\\[3pt]
                &\nu n\rightarrow l^{-}R^{+}~\\[3pt]
                &\nu p\rightarrow \nu R^{+}~\\[6pt]
                &\nu p\rightarrow \nu R^{0}~\\[3pt]
            \end{aligned}$
                        &
            $\begin{aligned}
            \\[-5pt]
                &F^p_i\\[3pt]
                &F^n_i\\[3pt]
                &-\\[3pt]
                &F^V_i=F^p_i - F_i^n\\[3pt]
                &\tilde{F}^p_i= s_WF^p_i - \frac{1}{2}F^n_i - \frac{1}{2}F^s_i\\[0.1pt]
                &\tilde{F}^p_i= s_WF^p_i - \frac{1}{2}F^n_i - \frac{1}{2}F^s_i\\[2pt]
            \end{aligned}$
                        &
         $\begin{aligned}
            \\[-5pt]
                &C^p_i\\[3pt]
                &C^n_i\\[3pt]
                &-\\[3pt]
                &C^V_i=C^p_i - C_i^n\\[3pt]
                &\tilde{C}^p_i= s_W~C^p_i - \frac{1}{2}C^n_i - \frac{1}{2}C^s_i\\[0.1pt]
                &\tilde{C}^p_i= s_W~C^p_i - \frac{1}{2}C^n_i - \frac{1}{2}C^s_i\\[3pt]
            \end{aligned}$
                        &
          $\begin{aligned}
            \\[-5pt]
                &F^N_i\\[3pt]
                &F^N_i\\[3pt]
                &F^V_i=-F^N_i\\[3pt]
                &F^V_i=-F^N_i\\[3pt]
                &\tilde{F}_i^N=s_W~F^N_i \\[4pt]
                &\tilde{F}_i^N=s_W~F^N_i\\[3pt] \end{aligned}$
                     &
          $\begin{aligned}
            \\[-5pt]
                &C^N_i\\[3pt]
                &C^N_i\\[3pt]
                &C^V_i=-C^N_i\\[3pt]
                &C^V_i=-C^N_i\\[3pt]
               &\tilde{C}_i^N=s_W~C^N_i \\[4pt]
               &\tilde{C}_i^N=s_W~C^N_i\\[3pt] \end{aligned}$                
        \end{tabular}
        \end{ruledtabular}
        \end{table*}
            
     The advanced theoretical framework developed in this work, referred to as the MK model, is designed to seamlessly incorporate these diverse datasets while remaining consistent with fundamental QCD principles and preserving all relevant symmetry relations. Such theoretical consistency is essential for achieving a unified and reliable description of nucleon dynamics across different experimental probes. 
    
    In addition, the analysis strategy employed in this study enables careful control of systematic uncertainties, which primarily originate from experimental inputs and model parametrisation choices. The resulting predictions include well-quantified uncertainties, a prerequisite for precision neutrino measurements and for reducing model-dependent biases. This capability is particularly important for next-generation neutrino experiments, where accurate interaction modelling is required to fully exploit the discovery potential of precision measurements.
        
\section{General description}
                
 Let us consider the weak, electro- and photo- pion production reactions:
                \begin{eqnarray}
                    \left.\begin{array}{l}
                         \nu_l  \\
                         \bar{\nu_l}
                    \end{array}\right\}
                    (k_1) + N(p_1)&\to&
                    \left.\begin{array}{c}
                         l  \\
                         \bar l
                    \end{array}\right\}
                    (k_2) + N(p_2) + \pi(q),\nonumber\\
                    e(k_1) + N(p_1)& \to& e(k_2) + N(p_2) + \pi(q),\nonumber\\
                    \gamma(k) + N(p_1)& \to&  N(p_2) + \pi(q)    
                \end{eqnarray}
                
                Here, $N$ denotes a nucleon, $\nu (\bar{\nu})$ a neutrino (anti-neutrino) and $l$ the outgoing charged lepton or neutrino in charged current (CC) and neutral current (NC) interactions, respectively. The four-momentum of each particle is indicated in parentheses. 
                
\begin{figure}[t]
                \centering
                \includegraphics [width=0.8\linewidth]{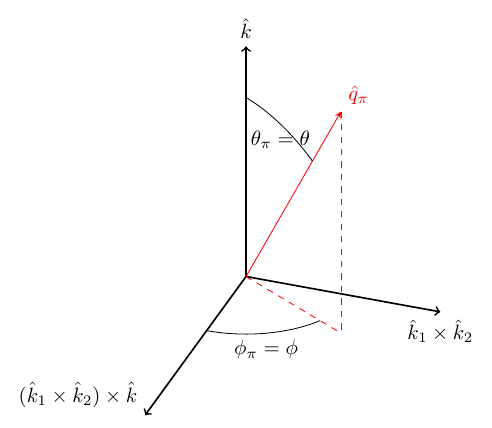}
                \caption{Isobaric (Adler) frame or the $\pi N $ center-of-mass frame. $\hat{k}$ is the direction of the momentum transfer between leptons.}\label{Isoframe}
                \end{figure}

                For weak and electroproduction processes, the four-momentum transfer between the initial and final leptons is defined as
                \begin{eqnarray}
                    k = k_1 - k_2,
                \end{eqnarray}
                 and the corresponding invariant four-momentum transfer squared is
                \begin{eqnarray}
                    Q^2 &=& -k^2 = -(k_1 - k_2)^2.
                \end{eqnarray}
                The invariant mass of the final pion-nucleon system, is 
                 \begin{eqnarray}
                W^2 &=& (k+p_1)^2 = (p_2 + q)^2.
                \end{eqnarray}
               For non-invariant quantities, such as energy transfer, a superscript $L$ is used to denote laboratory-frame variables. Quantities without the superscript $L$ refer to the center-of-mass frame of the final pion–nucleon system (the isobaric or Adler frame), which is defined by
        \begin{equation}
        \mathbf{q} + \mathbf{p}_2 = \mathbf{k} + \mathbf{p}_1 = 0 ,
        \end{equation}
        as illustrated in Fig.~\ref{Isoframe}.
    In this framework, the differential cross section for lepton-induced single-pion production is given by

    \begin{equation} \label{lep_xsec}
        \frac{d\sigma(l N  \rightarrow l'N\pi)}{dk^2 dW d\Omega_{\pi}} =
         \frac{1}{(2\pi)^4} \frac{|\bf{q}|}{8M^2} \frac{-k^2}{(\mathbf{k^L})^2}|\mathcal{M}|^2,
                \end{equation}
               
    while for photoproduction
\begin{equation} \label{photo_xsec}
\frac{d\sigma(\gamma N \rightarrow N \pi)}{d\Omega_{\pi}} =
\frac{1}{(4\pi)^2} \frac{2M^2}{W^2 - M^2}
\frac{|\mathbf{q}|}{W}
|\mathcal{M}|^2.
\end{equation}

The invariant matrix element can be written as
\begin{align}\label{matrix_element}
\mathcal{M}(\nu_l N \rightarrow l^- N \pi) &= \frac{G_F}{\sqrt{2}} a\, \epsilon_{\rho}\, \langle N\pi| J_{CC+}^{\rho} |N \rangle, \nonumber\\
\mathcal{M}(\bar{\nu}_l N \rightarrow l^+ N \pi) &= \frac{G_F}{\sqrt{2}} a\, \bar{\epsilon}_{\rho}\, \langle N\pi| J_{CC-}^{\rho} |N \rangle, \nonumber\\
\mathcal{M}(e N \rightarrow e' N \pi) &= e^2 \frac{1}{k^2} \epsilon_{\rho}^{EM}\, \langle N\pi| J_{EM}^{\rho} |N \rangle, \nonumber\\
\mathcal{M}(\gamma N \rightarrow N \pi) &= e\, \epsilon^{\gamma}_{\rho}\, \langle N\pi| J_{EM}^{\rho} |N \rangle.
\end{align}

Here $J^{\rho}$ denotes the hadronic current, $G_F$ is the Fermi constant, $e$ is the electric charge, and $a = \cos\theta_C$ for CC interactions and $a = 1/2$ for NC interactions. The photon polarization vector is denoted by $\epsilon^{\gamma}_{\rho}$, and the lepton currents are given by
\begin{eqnarray}
\epsilon_{\rho} &=& \bar{u}_l(k_2)\gamma_{\rho}(1-\gamma_5)u_{\nu}(k_1),\nonumber\\
\bar{\epsilon}_{\rho} &=& \bar{u}_{\bar{l}}(k_2)\gamma_{\rho}(1+\gamma_5)u_{\bar{\nu}}(k_1),\nonumber\\
\epsilon^{EM}_{\rho} &=& \bar{u}_e(k_2)\gamma_{\rho}u_e(k_1).
\label{epsilon}
\end{eqnarray}        
In the massless limit, only left-handed helicity states contribute. However, in charged-current interactions with a $\nu_{\mu}$ beam, the finite muon mass leads to non-negligible contributions from both helicity states. One can therefore define
 
\begin{eqnarray}
\epsilon^{\rho}_{\lambda} = \bar{u}_{l_{\lambda}}(k_2)\gamma_{\rho}(1-\gamma_5)u_{\nu}(k_1),
\end{eqnarray}
where $\lambda = -(+)$ corresponds to left- (right-)handed helicity.

The lepton current can be decomposed in terms of the polarization vectors of the intermediate boson,
\begin{eqnarray}
\epsilon^{\rho}_{\lambda} = C_{L_{\lambda}} e^{\rho}_{L} + C_{R_{\lambda}} e^{\rho}_{R} + C_{\lambda} e^{\rho}_{\lambda},
\end{eqnarray}
where $e^{\rho}_{L}$ and $e^{\rho}_{R}$ are transverse polarizations, and $e^{\rho}_{\lambda}$ is the longitudinal polarization. These are given by
\begin{eqnarray}
e^{\alpha}_{L} &=& \frac{1}{\sqrt{2}} (0,1,-i,0), \nonumber\\
e^{\alpha}_{R} &=& \frac{1}{\sqrt{2}} (0,-1,-i,0), \nonumber\\
e^{\alpha}_{\lambda} &=& \frac{1}{\sqrt{|(\epsilon^0_{\lambda})^2 - (\epsilon^3_{\lambda})^2|}} (\epsilon^0_{\lambda},0,0,\epsilon^3_{\lambda}),
\label{pol_basis}
\end{eqnarray}
with coefficients
\begin{eqnarray}
C_{L_{\lambda}} &=& \frac{1}{\sqrt{2}}(\epsilon_{\lambda}^{1}+ i\epsilon_{\lambda}^{2}), \nonumber\\
C_{R_{\lambda}} &=& -\frac{1}{\sqrt{2}}(\epsilon_{\lambda}^{1}- i\epsilon_{\lambda}^{2}), \nonumber\\
C_{\lambda} &=& \sqrt{|(\epsilon^0_{\lambda})^2 - (\epsilon^3_{\lambda})^2|}.
\label{lep_coeff}
\end{eqnarray}

The matrix element can be expressed in terms of helicity amplitudes and the coefficients in Eq.~\ref{lep_coeff}. The helicity amplitudes are defined by the helicities of the initial nucleon ($\lambda_1$), final nucleon ($\lambda_2$), and boson polarization ($\lambda_k$). This leads to 16 amplitudes for the vector current and 16 for the axial-vector current in CC interactions. For electromagnetic and NC processes, the number of independent amplitudes is reduced according to the available boson polarizations.

The helicity amplitudes are defined as
\begin{eqnarray}
\tilde{F}_{\lambda_2, \lambda_1}^{\lambda_k} &=& \langle N\pi|\, e^{\rho}_{\lambda_k} \frac{1}{2W} J^V_{\rho}\, |N \rangle, \nonumber\\
\tilde{G}_{\lambda_2, \lambda_1}^{\lambda_k} &=& \langle N\pi|\, e^{\rho}_{\lambda_k} \frac{1}{2W} J^A_{\rho}\, |N \rangle,
\end{eqnarray}
where $e^{\rho}_{\lambda_k}$ are the transverse and longitudinal polarization vectors of the virtual intermediate vector boson, as defined in Eq.~\ref{pol_basis}.

    The internal structure of the nucleon or excited nucleon are encoded in form factors \cite{Meyer:2016oeg,Meyer:2022mix,Meyer:2026kdl}, which depend on $Q^2$ and provide essential inputs to the hadronic currents and helicity amplitudes. The most sophisticated model to describe these form factors is the meson dominance (MD) model, based on the effective Lagrangian of quantum field theory as introduced in Refs.~\cite{Sakurai:1960ju, Gell-Mann:1961jim}. The MD model explains the interaction between leptons and nucleons through meson exchange with analogous quantum properties. The form factors can then be expressed in terms of the meson masses ($m_j$), the ratio of coupling strengths between the gauge boson and the meson, and between the meson and the nucleon ($a_j$), summing over all possible $n$ mesons:              
        \begin{equation}
            F(k^2)= \sum_{j=1}^n \frac{a_j m^2_j}{m_j^2 - k^2}
                \label{FFN}
        \end{equation}
                where a list of vector-mesons and axial-mesons is given in Table~\ref{vmeson}.
                 \begin{table}
                    \centering
                    \caption{(axial-) vector meson masses}
                    \label{vmeson}
                    \renewcommand{\arraystretch}{1.3}
                    \begin{ruledtabular}
                    \begin{tabular}[t]{lcccl}
                        k&$ \rho$ -group  & $m_{(\rho)k}[\text{GeV}]$ &$\omega$ -group&  $m_{(\omega)k}[\text{GeV}]$ 
                        \\ [0.1ex]
                        \hline
                        1&$\rho(770)$ & 0.77526 & $\omega(782)$& 0.78265\\
                        2&$\rho(1450)$ & 1.465 & $\omega(1420)$& 1.410\\
                        3&$\rho(1700)$ & 1.720 & $\omega(1650)$& 1.670\\
                        4&$\rho(1900)$ & 1.885 & $\omega(1960)$& 1.960\\
                    \hline
                    \vspace{0.1cm}
                        k&$a_1$ -group  & $m_{(a_1)k}[\text{GeV}]$ &$f_1$ -group&  $m_{(f_1)k}[\text{GeV}]$ 
                              \\ [0.1ex]
                        \hline
                        1&$a_1(1260)$ & 1.230 & $f_1(1285)$& 1.2819\\
                        2&$a_1(1420)$ & 1.411 & $f_1(1420)$& 1.4263\\
                        3&$a_1(1640)$ & 1.655 & $f_1(1510)$& 1.518\\
                        4&$a_1(2095)$ & 2.096 & $f_1(1970)$& 1.1971\\              
                    \end{tabular}
                    \end{ruledtabular}
                    \label{meson_mass}
                    \end{table}       
                
                There are two shortcoming from the form factor model; firstly, the form factor in Eq.~\ref{FFN} does not have analytic properties and does not obey the unitarity condition. Secondly, the asymptotic behaviour of the form factor differs from the asymptotic scaling behaviours predicted by QCD. However these properties can be imposed to the MD model.
                
                 Imposing the conditions bring us to a complex system of equations for the coupling constant ratios, making it difficult to find solutions in the general case. However, in Ref. \cite{Adamuscin:2002ca}, an equivalent system of equations is derived for the coupling constant ratios, with coefficients that are simply even powers of the corresponding meson masses. In principle, one can find solutions to these equations even in the general case, referred to as linear superconvergence relations:
                \begin{align}\label{sup_converged}
\sum_{j=1}^n a_j &=1,\nonumber\\                    \sum_{j=1}^n m^2_j a_j &=0,\nonumber\\
                    \sum_{j=1}^n m^4_j a_j &=0,\nonumber\\
                   &\vdots\nonumber\\
                   \sum_{j=1}^n m^{2(m-1)}_j a_j &=0,
                \end{align}
                where $m\le n$ shows the asymptotic behaviour of the form factors:
                \begin{equation*}
                   F(k^2)_{|k^2| \rightarrow \infty} \sim k^{-2m} 
                \end{equation*}
        \section{Resonance production}
    
    The general form of the helicity amplitudes was derived in our previous works~\cite{MK,MK2,MK3} and is reproduced in Appendix~\ref{appA} for completeness. The sixteen vector helicity amplitudes ($\tilde{F}_{\lambda_2, \lambda_1}^{\lambda_k}$) are presented in Table~\ref{res_HV}, while the sixteen axial-vector helicity amplitudes ($\tilde{G}_{\lambda_2, \lambda_1}^{\lambda_k}$) are listed in Table~\ref{res_HAA}. These amplitudes incorporate several ingredients, including Clebsch-Gordan coefficients corresponding to different SPP channels and Breit-Wigner parametrisations describing the finite widths of the resonance states.

The key element in their definition is the resonance production amplitude, which is the focus of this section. These amplitudes are defined as                
            \begin{eqnarray}\label{HAmp}
                f^{(V,A)}_{-3} &=& -\frac{1}{2W}\langle~N^*, 3/2|~e^{\rho}_{R}  J^{(V,A)}_{\rho}~|N, 1/2 ~\rangle \nonumber\\
                    f^{(V,A)}_{-1} &=& -\frac{1}{2W}\langle~N^*, 1/2|~e^{\rho}_{R}  J^{(V,A)}_{\rho}~|N, -1/2 ~\rangle \nonumber\\
                    f^{(V,A)(\lambda)}_{0+} &=& \frac{1}{2W}\frac{\sqrt{-k^2}}{|{\bf{k}}|}\langle~N^*, 1/2|~e^{\rho}_{R}  J^{(V,A)}_{\rho}~|N, 1/2 ~\rangle \nonumber\\
            \end{eqnarray}
Here $f^{(V,A)}$ denote the vector and axial-vector production amplitudes, and $J^{(V,A)}_\rho$ represent the corresponding hadronic currents. The state $N^*$ denotes the produced resonance.
                           
    The resonances in the first and second resonance regions have spin $1/2$ or $3/2$. The corresponding hadronic currents associated with these resonances can be written as
    
        \begin{eqnarray}
            \langle~N^*(p)|~J^{\rho}_{3/2}~|N(p_1) ~\rangle &=& \bar{\psi}_{\alpha}(p)\Gamma^{\alpha \rho}_{3/2}u(p_1)\nonumber\\
            \langle~N^*(p)|~J^{\rho}_{1/2}~|N(p_1) ~\rangle &=& \bar{u}(p)\Gamma^{\rho}_{1/2}u(p_1) ,  \nonumber\\ 
        \end{eqnarray}   
                  
        where $p=p_1 + k$. In these equations, $u(p)$ represents the Dirac spinor for spin-1/2 resonances, while $\psi_{\alpha}(p)$ denotes the Rarita–Schwinger spinor for spin-3/2 resonances \cite{Lalakulich:2006sw}. The operators $\Gamma^{\alpha \rho}_{3/2}$ and $\Gamma^{\rho}_{1/2}$ encapsulate the interaction dynamics and depend on the specific form of the current $J^{\rho}$, as well as the quantum numbers of the resonances involved. Given that the dynamics governing resonances with spin 1/2 and 3/2 differ significantly, I will discuss these cases separately to provide a clear understanding of their respective contributions:
                \subsection{Resonance with spin $1/2$ }
                In the first and second resonance regions, there are two resonances with spin $1/2$: the $P_{11}(1440)$ resonance, which has positive parity, and the $S_{11}(1535)$ resonance, which has negative parity. Their corresponding interactions are defined as follows:
            
                \begin{eqnarray}
                    \Gamma^{\rho}_{1/2}(P_{11}) &=& \left[\mathcal{V}^{ \rho}_{1/2} - \mathcal{A}^{ \rho}_{1/2}\right]\nonumber\\
                    \Gamma^{\alpha \rho}_{1/2}(S_{11}) &=& \left[\mathcal{V}^{\alpha \rho}_{1/2} - \mathcal{A}^{\alpha \rho}_{1/2}\right]\gamma^5
                \end{eqnarray} 
                where the vector ($\mathcal{V}^{\alpha \rho}_{1/2}$) and axial-vector ($\mathcal{A}^{\alpha \rho}_{1/2}$) components are given by:
                \begin{eqnarray}\label{Gamma1}
                    \mathcal{V}^{\rho}_{1/2} &=& \frac{\mathcal{F}_1^V}{2M^2}\left(\not{k}k^{\rho} - k^2 \gamma^{\rho} \right) + \frac{\mathcal{F}_2^V}{2M}\left(i\sigma^{\rho \alpha}k_{\alpha} \right)~,\nonumber\\
                    \mathcal{A}^{\rho}_{1/2} &=& \mathcal{F}_A\gamma^{\rho}\gamma^5  + \frac{\mathcal{F}_P}{M} k^{\rho} \gamma^5~.
                \end{eqnarray}
                The vector and axial-vector helicity amplitudes for the $P_{11}$ and $S_{11}$ resonances can be derived from Eqs.~\ref{HAmp} and \ref{Gamma1}. Here, I present the explicit forms of these amplitudes for $P_{11}$ resonance:
                  \begin{eqnarray}
                  f^{V}_{-1} (P_{11})&= &\frac{|{\bf{k}}|}{\sqrt{W(E_k +M)}} \left[\mathcal{F}_1^V - \frac{\mathcal{F}_2^V}{W_{+}^2} k^2\right] \nonumber\\
                f^{V(\lambda)}_{0+} (P_{11})&= &-\frac{1}{C_{\lambda}}\frac{|{\bf{k}}|}{\sqrt{2W(E_k +M)} } \left(|{\bf{k}}|\epsilon^0_{\lambda} - k_0\epsilon^3_{\lambda} \right) \nonumber\\
                &~&~~\frac{1}{W_{+}}\left[ \mathcal{F}_1^V - \mathcal{F}_2^V\right]
                 \end{eqnarray} 
                 \begin{eqnarray}
                 f^{A}_{-1}(p_{11}) &= & - \mathcal{F}_A \sqrt{\frac{E_k + M}{W}}\nonumber\\
                f^{A(\lambda)}_{0+} (P_{11})&= &\frac{1}{C_{\lambda}}\frac{1}{\sqrt{2W(E_k +M)} } \bigg [ \mathcal{F}_A\big [\left(|{\bf{k}}|\epsilon^0_{\lambda} - k_0\epsilon^3_{\lambda} \right) \nonumber\\
                &~& ~+ \epsilon^3_{\lambda}W_{+}\big ] + \mathcal{F}_P\frac{|{\bf{k}}|}{M}\left(|{\bf{k}}\epsilon^0_{\lambda} - k_0\epsilon^3_{\lambda} \right) \bigg ]
                \end{eqnarray} 
                and for $S_{11}$ resonance:        
                \begin{eqnarray}
                  f^{V}_{-1} (S_{11})&= &\sqrt{\frac{(E_k + M)}{W}}\left[\frac{\mathcal{F}_1^V}{W_{+}^2} k^2 - \frac{\mathcal{F}_2^V}{W_{+}}W_{-} \right] \nonumber\\
                f^{V(\lambda)}_{0+} (S_{11})&= &\frac{1}{C^V_{\lambda}} \sqrt{\frac{E_k + M}{2W}}\left(|{\bf{k}}|\epsilon^0_{\lambda} - k_0\epsilon^3_{\lambda} \right)\nonumber\\
                &~&\left[-\frac{\mathcal{F}_1^V W_{-}}{W_{+}^2} + \frac{\mathcal{F}_2^V}{W_{+}}\right] 
                \end{eqnarray} 
                 \begin{eqnarray}
                 f^{A}_{-1} (S_{11})&= &-\mathcal{F}_A \frac{|{\bf{k}}|}{\sqrt{W(E_k +M)}}\nonumber\\
                     f^{A(\lambda)}_{0+}(S_{11}) &= &-\frac{1}{C_{\lambda}}\sqrt{\frac{E_k +M} {3W}} \bigg[  \mathcal{F}_A\left( \epsilon^0_{\lambda}W_{+}\right)\nonumber\\ &~& + \frac{\mathcal{F}_P}{M}\left(|{\bf{k}}|\epsilon^0_{\lambda} - k_0\epsilon^3_{\lambda} \right) \bigg]
                 \end{eqnarray} 
                 
            where $\mathcal{F}_i^V$ ($i=1,2$) denote the vector transition form factors for spin-$1/2$ resonances and correspond to either the charged-current (CC) form factors ($F_i^V$), the electromagnetic form factors ($F_i^N$, with $N=p,n$), or the neutral-current (NC) form factors ($\tilde{F}_i^N$, with $N=p,n$). Similarly, $\mathcal{F}_A$ and $\mathcal{F}_P$ denote the CC axial-vector and pseudoscalar form factors ($F_A$, $F_P$) or their NC counterparts ($\tilde{F}_A^N$, $\tilde{F}_P^N$). The pseudoscalar form factor $F_P$ is not independent, but is related to $F_A$ through the PCAC hypothesis~\cite{Lalakulich:2006sw,HNV}. Consequently, only one independent axial-vector form factor is required for spin-$1/2$ resonances.
                        
            The vector form factors $C_i^V$ and $C_i^{VN}$ ($F_i^V$ and $F_i^{VN}$) can be related to the electromagnetic form factors $C_i^{p,n}$ ($F_i^{p,n}$) as is presented in Table \ref{Isospin_symmetry}.   
          
    \subsection{Resonances with spin $3/2$ }
                In the first and second resonance regions, there are two resonances with spin $3/2$: the $P_{33}$ ($\Delta$) resonance, which has positive parity, and the $D_{13}(1520)$ resonance, which has negative parity. Their corresponding interactions are defined as follows:
                \begin{eqnarray}
                    \Gamma^{\alpha \rho}_{3/2}(P_{33}) &=& \left[\mathcal{V}^{\alpha \rho}_{3/2} - \mathcal{A}^{\alpha \rho}_{3/2}\right]\gamma^5\nonumber\\
                    \Gamma^{\alpha \rho}_{3/2}(D_{13}) &=& \left[\mathcal{V}^{\alpha \rho}_{3/2} - \mathcal{A}^{\alpha \rho}_{3/2}\right]
                \end{eqnarray} \label{Gam}
                where the vector ($\mathcal{V}^{\alpha \rho}_{3/2}$) and axial-vector ($\mathcal{A}^{\alpha \rho}_{3/2}$) components are given by:
                \begin{eqnarray}\label{Gamma}
                    \mathcal{V}^{\alpha \rho}_{3/2} &=& \frac{\mathcal{C}_3^V}{M}\left(g^{\alpha \rho}\not{k} - k^{\alpha} \gamma^{\rho} \right) + \frac{\mathcal{C}_4^V}{M^2}\left(g^{\alpha \rho} k.p - k^{\alpha} p^{\rho} \right) \nonumber\\
                    &~& + \frac{\mathcal{C}_5^V}{M^2}\left(g^{\alpha \rho}k.p_1 - k^{\alpha} p_1^{\rho} \right)~,\nonumber\\
                    \mathcal{A}^{\alpha \rho}_{3/2} &=& \frac{\mathcal{C}_3^A}{M}\left(g^{\alpha \rho}\not{k} - k^{\alpha} \gamma^{\rho} \right) + \frac{\mathcal{C}_4^A}{M^2}\left(g^{\alpha \rho} k.p - k^{\alpha} p^{\rho} \right)\nonumber\\ &~& + \mathcal{C}_5^A g^{\alpha \rho} + \frac{\mathcal{C}_6^A}{M^2}k^{\alpha} k^{\rho}~.
                \end{eqnarray}
                
                The vector and axial-vector helicity amplitudes for the $P_{33}$ and $D_{13}$ resonances can be derived from Eqs.~\ref{HAmp} and \ref{Gamma}. Here, I present the explicit forms of these amplitudes for $P_{33}$ resonance:
            
        \begin{widetext}
            \begin{eqnarray}
                  f^{V}_{-3} (P_{33})&= &-\frac{|{\bf{k}}|}{W}\frac{1}{\sqrt{2\mathcal{N}} }\left[\frac{\mathcal{C}_3^V W_{+}}{M} 
                  + \frac{\mathcal{C}_4^V}{M^2} Wk_0 +  \frac{\mathcal{C}_5^V}{M^2} (Wk_0 -k^2 )\right],\nonumber\\
                 f^{V}_{-1} (P_{33})&= &\frac{|{\bf{k}}|}{W}\frac{1}{\sqrt{2\mathcal{N}} }\left[\frac{\mathcal{C}_3^V}{MW}\left(k^2 - MW_{+}\right) 
                  + \frac{\mathcal{C}_4^V}{M^2} Wk_0 +  \frac{\mathcal{C}_5^V}{M^2} (Wk_0 -k^2)\right],\nonumber\\
                f^{V(\lambda)}_{0+}(P_{33}) &= &-\frac{|{\bf{k}}|}{W}\frac{1}{\sqrt{2\mathcal{N}} }\frac{1}{\mathcal{C}_{\lambda}} \left(|{\bf{k}}|\epsilon^0_{\lambda} - k_0\epsilon^z_{\lambda} \right) \left[\frac{\mathcal{C}_3^V}{M}  + \frac{\mathcal{C}_4^V}{M^2}W  +  \frac{\mathcal{C}_5^V}{M^2} (W- k_0)\right],
            \end{eqnarray} 
                 \begin{eqnarray}
                  f^{A}_{-3}(P_{33}) &= &- \sqrt{\frac{\mathcal{N}}{2}}\left [\frac{\mathcal{C}_3^A W_{-}}{M} + \frac{\mathcal{C}_4^A}{M^2} Wk_0 +  \mathcal{C}_5^A \right],\nonumber\\
                   f^{A}_{-1} (P_{33})&= &- \sqrt{\frac{\mathcal{N}}{6}}\left [\frac{\mathcal{C}_3^A}{M} \frac{2|{\bf{k}}| + k_0W_{-}}{E_k + M} - \frac{\mathcal{C}_4^A}{M^2} Wk_0 -  \mathcal{C}_5^A \right],\nonumber\\
                f^{A(\lambda)}_{0+}(P_{33}) &= &-\sqrt{\frac{\mathcal{N}} {3}} \frac{1}{\mathcal{C}_{\lambda}}\left[\left(\frac{\mathcal{C}_3^A}{M}  + \frac{\mathcal{C}_4^A}{M^2}W\right)\left(|{\bf{k}}|\epsilon^0_{\lambda} - k_0\epsilon^z_{\lambda} \right)   -  \mathcal{C}_5^A \epsilon^z_{\lambda} + \frac{\mathcal{C}_6^A}{M^2} |{\bf{k}}|\left(|{\bf{k}}|\epsilon^0_{\lambda} - k_0\epsilon^z_{\lambda} \right) \right],
                 \end{eqnarray} 
                 and for $D_{13}$ resonance:
                 \begin{eqnarray}
                  f^{V}_{-3} (D_{13})&= &\sqrt{\frac{\mathcal{N}}{2}}\left[\frac{\mathcal{C}_3^V}{M}W_{-} + \frac{\mathcal{C}_4^V}{M^2} Wk_0
                  +  \frac{\mathcal{C}_5^V}{M^2} \left(Wk_0 - k^2\right)\right],\nonumber\\
                   f^{V}_{-1} (D_{13})&= &\sqrt{\frac{\mathcal{N}}{6}}\left[\frac{\mathcal{C}_3^V}{M}\left(W _{-}- 2\frac{{\bf{k}}^2}{E_k + M} \right) 
                 + \frac{\mathcal{C}_4^V}{M^2}Wk_0  +  \frac{\mathcal{C}_5^V}{M^2}\left(Wk_0 - k^2\right)  \right],\nonumber\\
                f^{V(\lambda)}_{0+} (D_{13})&= &\sqrt{\frac{\mathcal{N}}{3}}\frac{1}{\mathcal{C}_{\lambda}} \left(|{\bf{k}}|\epsilon^0_{\lambda} - k_0\epsilon^z_{\lambda} \right) \left[\frac{\mathcal{C}_3^V}{M}  + \frac{\mathcal{C}_4^V}{M^2}W  +   \frac{\mathcal{C}_5^V}{M^2} (W- k_0)\right] ,
                \end{eqnarray} 
                \begin{eqnarray}
                  f^{A}_{-3} (D_{13})&= &-\frac{|{\bf{k}}|}{W}\frac{1}{\sqrt{2\mathcal{N}} }\left[\frac{\mathcal{C}_3^A W_{+}}{M} + \frac{\mathcal{C}_4^A}{M^2} Wk_0 +  \mathcal{C}_5^A \right],\nonumber\\
                 f^{A}_{-1} (D_{13})&= & \frac{|{\bf{k}}|}{W}\frac{1}{\sqrt{6\mathcal{N}} }\left [\frac{\mathcal{C}_3^A}{M} (W+M-2k_0) - \frac{\mathcal{C}_4^A}{M^2} Wk_0 -  \mathcal{C}_5^A \right],\nonumber\\
                f^{A(\lambda)}_{0+} (D_{13})&= & \frac{|{\bf{k}}|}{W}\frac{1}{\sqrt{3\mathcal{N}} }\frac{1}{\mathcal{C}_{\lambda}} \left[\left(\frac{\mathcal{C}_3^A}{M}  + \frac{\mathcal{C}_4^A}{M^2}W\right)\left(|{\bf{k}}|\epsilon^0_{\lambda} - k_0\epsilon^z_{\lambda} \right)   -  \mathcal{C}_5^A \epsilon^z_{\lambda} + \frac{\mathcal{C}_6^A}{M^2} |{\bf{k}}|\left(|{\bf{k}}|\epsilon^0_{\lambda} - k_0\epsilon^3_{\lambda} \right) \right] ,
                \end{eqnarray} 
                \end{widetext}
                where $W\pm=W \pm M$, $E_k=\sqrt{k^2 + M^2}$ and $\mathcal{N}=\frac{E_k + M}{W}$.

    The coupling factors $\mathcal{C}_i^V$ ($i=3$--$5$) denote the vector transition form factors for spin-$3/2$ resonances and correspond to either the charged-current (CC) form factors ($C_i^V$), the electromagnetic form factors ($C_i^N$, with $N=p,n$), or the neutral-current (NC) form factors ($\tilde{C}_i^N$, with $N=p,n$). Similarly, the coupling factors $\mathcal{C}_i^A$ ($i=3$--$6$) represent the CC axial-vector form factors ($C_i^A$) or the NC axial-vector form factors ($\tilde{C}_i^A$). The pseudoscalar form factor $C_6^A$ is related to $C_5^A$ through the partially conserved axial current (PCAC) hypothesis~\cite{Lalakulich:2006sw,HNV}. Consequently, only three independent axial-vector form factors are required.
   
            \subsection{Resonances with higher spin}
        The third resonance region is characterised by overlapping contributions from multiple baryon resonances. Among these, only two states, namely the $F_{15}(1680)$ and $F_{37}(1950)$, give dominant contributions to the hadronic tensor in the regions around $W \simeq 1.6$ GeV and $W \simeq 1.9$ GeV, respectively. Both resonances have spins greater than $3/2$, which significantly increases the complexity of their theoretical description due to the increased number of degrees of freedom.
        
        To avoid introducing an excessive number of free parameters and to maintain control over model uncertainties in this densely populated region, only these two resonances are retained in the present description of the third resonance region. For simplicity, they are treated using the spin $3/2$ formalism described earlier in this section. This approach keeps the model tractable while still capturing the essential physics of the third resonance region.
        
\section{ Form Factor Parametrisation} \label{FF}

The MD parametrisation employed in the MK model provides an effective description of nucleon and excited-nucleon transition form factors across both the perturbative and non-perturbative regimes, consistent with QCD expectations, as discussed in Ref.~\cite{stoler}. In particular, it provides a suitable framework for describing the hadron-quark transition region. Throughout this work, the parametrisation proposed in Ref.~\cite{Vereshkov:2007xy} is adopted. 

For spin-$1/2$ resonances, the vector and axial-vector transition form factors are parametrised as 
    \begin{eqnarray} \label{FFCg}
    F_i^{(V,A)}(Q^2)&=& \frac{F_i^{(V,A)}(0)}{L_i(Q^2)}\sum_{k=1}^{k=4} \frac{a^{(V,A)}_{i k }m^2_{k(V,A)}}{m_{k(V,A)}^{2}+ Q^2}, ~~(i=1-2)  \nonumber\\
\end{eqnarray} 

where $F_i^{V}(Q^2)$ ($F_i^{A}(Q^2)$) denote the vector (axial-vector) transition form factors and $F_i^{V(A)}(0)$  are their values at $Q^2=0$ determined from the fit, and $m_V = \frac{m_{\omega} + m_{\rho}}{2}$ ($m_A=\frac{m_{a_1} + m_{f_1}}{2}$) where the meson masses given in Table~\ref{meson_mass}. The axial form factors $F_1^{A}$ and $F_2^{A}$ correspond to the axial-vector and pseudoscalar form factors, $F_A$ and $F_P$, respectively, introduced in the previous section.
    
    For spin-$3/2$ resonances, the transition form factors are written as
\begin{eqnarray} \label{FFCg}
    C_i^{(V,A)}(Q^2)&=& \frac{C_i^{(V,A)}(0)}{L_i(Q^2)}\sum_{k=1}^{k=4} \frac{a^{(V,A)}_{i k }m^2_{k(V,A)}}{m_{k(V,A)}^2+ Q^2}, ~~(i=3-5)  \nonumber\\
\end{eqnarray} 

where $C_i^{V}(Q^2)$ and $C_i^{A}(Q^2)$ denote the vector and axial-vector transition form factors, respectively, with the corresponding masses given by $m_V=m_\rho$ and $m_A=m_{a_1}$, and $C_i^{V,A}(0)$ are their values at $Q^2=0$.
        
    The parameters $a^{V(A)}_{ i k }$ which represent the coupling constant ratios of vector (axial-vector) mesons to nucleons, are not free parameters. Instead, they are constrained by the linear superconvergence relations as expressed in Eq.~\ref{sup_converged}:

                 For $i=1$ and $3$:
                        \begin{eqnarray} 
                          &\sum_{k=1}^{K}& a_{i k }=1,~~~~~~~~~~~~~~~\sum_{k=1}^{K} a_{i k } m^2_{k}=0,\nonumber\\ 
                          &\sum_{k=1}^{K}& a_{i k } m^4_{k}=0.
                        \end{eqnarray} 
                        
                        For $i=2,4$ and $5$:
                        \begin{eqnarray} 
                          &\sum_{k=1}^{K}& a_{i k }=1,~~~~~~~~~~~~~~\sum_{k=1}^{K} a_{i k } m^2_{k}=0,\nonumber\\ 
                          &\sum_{k=1}^{K}& a_{i k } m^4_{k}=0,~~~~~~~~~~\sum_{k=1}^{K} a_{i k }m^6_{k}=0.
                        \end{eqnarray}

        The logarithmic renormalisation $L_{i}(k^2)$ is required to incorporate QCD quark subprocess contributions in the asymptotic region, thereby allowing the MD model’s predictions to match perturbative QCD expectations. In Ref. \cite{Vereshkov:2007xy}, only the leading contributions were included. In this work, I also include nonleading effects inspired by the higher twist expansion \cite{SajjadAthar:2022pjt, Jorge_SIS} in order to describe the transition region. 
    \begin{eqnarray} \label{log}
        L_{i}(Q^2) = 1 +  k_{i}\ln^{n_i}\left(1 + \frac{Q^2}{\Lambda^2_{QCD}}\right)  
    \end{eqnarray} 
        where $\Lambda_{QCD}\in [0.19-0.24]$ is QCD scale and
    \begin{eqnarray} \label{twist}
       k_{i}= k_{i_{0}} + \frac{k_{i_{2}}}{Q^2} + \frac{k_{i_{4}}}{Q^4}+ \cdots .
    \end{eqnarray} 
        $k_{i_{j}},~ j=0,2,4,\cdots$ are free parameters of the model and $n_1 = n_3 \simeq 3$, $n_1>n_2$ and $n_5>n_3>n_4$.
    
        The meson masses entering the MD parametrisation have finite experimental widths, which introduce an intrinsic uncertainty in the form factors. Such effects can be estimated through the half-width rule proposed in Ref.~\cite{Masjuan:2012sk}, and similar ideas have been applied to the axial form factor in quasielastic neutrino scattering~\cite{Amaro:2015lga}. In the present work, however, the meson-pole contributions are accompanied by fitted coefficients $a_{ik}^{(V,A)}$ and logarithmic factors $L_i(Q^2)$, whose parameters are determined simultaneously in the global analysis. Since these quantities are strongly correlated, the finite-width uncertainty of the meson poles is not treated as an independent source of uncertainty. The quoted uncertainties are obtained from the full covariance matrix of the fit, which consistently propagates the correlations among all fitted parameters.    
        
\section{Non-resonant background} 
    
    The treatment of non-resonant interactions follows the framework developed in my previous work~\cite{MK3}, which is based on the Hybrid model~\cite{Gonzalez-Jimenez:2016qqq}. This approach extends the range of validity of chiral perturbation theory~\cite{HNV} to higher invariant masses $W$ through the inclusion of a Regge-trajectory description~\cite{guidal:1997hy}. The form factors entering the helicity amplitudes, however, differ from those used previously. In the present work, the nucleon form factors are treated analogously to those of spin-$1/2$ resonances, following the parametrisation introduced in Ref.~\cite{Vereshkov:2007um}. As a result, the non-resonant contribution employed here remains applicable over a broad range of $Q^2$ and $W$.
    
    The vector and axial-vector helicity amplitudes for the non-resonant background were derived in my earlier work~\cite{MK,MK2}.

        \section{Analysis of exclusive SPP data}
        
        Accurate determination of the vector and axial vector helicity amplitudes for both resonant and non-resonant interactions is essential for providing reliable predictions of SPP channels. These amplitudes depend on several parameters, some of which can be constrained by symmetry relations and fundamental principles, while others can only be determined using experimental data.
        
        In this section, I present a comprehensive analysis of exclusive SPP cross-section measurements on hydrogen and deuterium targets to determine the $Q^2$ dependence of the proton, neutron, and axial form factors described in the previous section. This study provides a detailed understanding of the underlying interaction dynamics. The analysis incorporates a wide range of data sets spanning a broad kinematic region to ensure robust and accurate form factor extractions, while accounting for differences in experimental conditions and measurement techniques. The primary goal is to extract these form factors and to quantify the associated systematic uncertainties across the full kinematic range.
        
        The vector form factors entering weak interactions are closely related to electromagnetic form factors as presented in Table \ref{Isospin_symmetry}. As a result, electron and photon scattering measurements on nucleon targets provide strong constraints on the vector sector. In contrast, pion and neutrino scattering data are essential for determining the axial vector form factors. By exploiting all available cross-section data within a combined analysis, the unified MK model achieves the most precise predictions for SPP channels currently possible. This integrated approach ensures reliable coverage of the full kinematic range relevant for SPP processes.

        \subsection {Photo-Nucleon and Electro-Nucleon Single Pion Productions}

        Electron- and photon-induced single-pion production provide the most direct constraints on the vector transition form factors entering the weak interaction current. The availability of monochromatic incident beams together with the broad kinematic coverage of modern measurements makes these reactions particularly well suited for determining the $Q^2$ dependence of the resonance transition form factors. Since theoretical calculations do not reliably predict this dependence throughout the resonance and transition regions, the form factors are determined phenomenologically through a global fit to the available electromagnetic scattering data.

The SPP channels for electron and photon interactions with protons are

            \begin{eqnarray}
                    e + p \rightarrow e p ~\pi^{0}~~&,&~~~~~~~~~ e + p \rightarrow  e n ~\pi^{+} \label{ep}\\
                    \gamma + p \rightarrow  p ~\pi^{0}~~&,&~~~~~~~~~
                \gamma + p \rightarrow  n ~\pi^{+} \label{gammap}
            \end{eqnarray}       
    Similarly, the corresponding reactions on neutrons are
    
            \begin{eqnarray}
                e + n \rightarrow  e n ~\pi^{0}~~ &,& ~~~~~~~~~ 
                e + n \rightarrow  e p ~\pi^{-}~~ \nonumber\\
                \gamma + n \rightarrow  n ~\pi^{0}~~ &,& ~~~~~~~~~ 
                \gamma + n \rightarrow  p ~\pi^{-}. \label{en}
            \end{eqnarray}

        In previous work presented in Ref. \cite{MK3}, the MD parametrisation was developed and tested using electron-proton scattering data in the two channels of Eq.~(\ref{ep}). The corresponding form factors were extracted from electroproduction measurements over a broad kinematic range, demonstrating that the model provides a smooth and accurate description of the available data throughout the resonance region.
        
        The present work significantly extends that analysis in several important ways. First, photo-proton scattering data, Eq.~(\ref{gammap}), are incorporated into the global fit. These measurements provide direct constraints on the vector form factors at the photon point ($Q^2=0$), complementing the electroproduction data and improving the determination of the low-$Q^2$ behaviour of the transition form factors. Second, recent measurements of $\pi^-$ electroproduction~\cite{CLAS:2022kta} make it possible, for the first time, to perform a simultaneous analysis of electron- and photon-induced reactions on neutron targets, Eq.~(\ref{en}). This enables the extraction of the electromagnetic transition form factors of excited neutron states, which have been considerably less constrained experimentally than their proton counterparts.

A summary of the experimental data included in the analysis is given in Table~\ref{clas_data}. The data span a broad kinematic region,
\[
Q^2 = 0.0\text{--}6.0~\mathrm{GeV}^2,
\qquad
W = 1.1\text{--}2.01~\mathrm{GeV},
\]
providing comprehensive coverage of the resonance and transition regions.

The analysis is based exclusively on differential cross-section measurements with published systematic uncertainties. For photoproduction, the complete differential cross-section database compiled by the SAID collaboration~\cite{Workman:2012jf} is employed. The measured differential cross sections correspond directly to the theoretical cross section defined in Eq.~\ref{photo_xsec}.

For electron-proton scattering, the complete set of CLAS cross-section measurements~\cite{database_CLAS} is used. The present analysis is restricted to unpolarized differential cross sections, which provide the most comprehensive and internally consistent dataset over the full kinematic region considered in this work.

The available electroproduction measurements include

\[
\frac{d\sigma_T}{d\Omega_\pi^\ast}
+\epsilon
\frac{d\sigma_L}{d\Omega_\pi^\ast},
\qquad
\frac{d\sigma_{LT}}{d\Omega_\pi^\ast},
\qquad
\frac{d\sigma_{TT}}{d\Omega_\pi^\ast},
\qquad
\frac{d\sigma_{LT'}}{d\Omega_\pi^\ast}.
\]

These observables are related to the fully differential cross section for lepton scattering in Eq. \ref{lep_xsec}, through

 \vbox{\begin{align}
     &\frac{d \sigma}{dE_{e'} d\Omega_{e'} d\Omega^{\ast} _{\pi}}= \frac{MEE_{e'}}{\pi W}\frac{d \sigma}{dk^2 dW d\Omega _{\pi}}\nonumber\\
    & =\Gamma_{em} \Big[ \frac{d\sigma_T}{d\Omega^{\ast} _{\pi} }   + \epsilon \frac{d\sigma_L}{d\Omega^{\ast} _{\pi} }  + \sqrt{2\epsilon(1+ \epsilon)} \frac{d\sigma_{LT}}{d\Omega^{\ast} _{\pi} } \cos \phi^{\ast}_{\pi}\nonumber\\
  & + \epsilon \frac{d\sigma_{TT}}{d\Omega^{\ast} _{\pi} }\cos 2\phi^{\ast}_{\pi}  + h_e \sqrt{2\epsilon(1+ \epsilon)} \frac{d\sigma_{LT'}}{d\Omega^{\ast} _{\pi} } \sin \phi^{\ast}_{\pi}  \Big]~,
 \label{em_Xsec}
  \end{align}}  
  
where $E_e$ and $E_{e'}$ are the energies of the incoming and scattered electrons in the laboratory frame, $\Omega_\pi^\ast=\Omega_\pi$ denotes the pion solid angle in the hadronic centre-of-mass frame, and
\begin{equation}
\Gamma_{em}
=
\frac{\alpha}{2\pi^2Q^2}
\frac{E_e}{E_{e'}}
\frac{q_\gamma}{1-\epsilon},
\end{equation}

is the virtual-photon flux factor with
\begin{equation}
    q_\gamma=\frac{W^2-M^2}{2M},
\end{equation}

while
\[
\epsilon=
\left[
1+
2\frac{q_\gamma^2}{Q^2}
\tan^2\left(\frac{\delta}{2}\right)
\right]^{-1}
\]
is the virtual-photon polarization parameter and $\delta$ is the electron scattering angle. The electron-neutron coss-section measurements are 
are related to the fully differential lepton scattering cross section (Eq.~\ref{lep_xsec}) by
\begin{equation}
\frac{d\sigma}{d\Omega_\pi}
=
\frac{1}{\Gamma_{em}}
\frac{d\sigma}{dQ^2\,dW\,d\Omega_\pi}.
\end{equation}

The transition form factors extracted from proton and neutron targets are related by isospin symmetry. For isospin-$3/2$ resonances, the proton and neutron transition form factors are identical, whereas for isospin-$1/2$ resonances they are independent quantities. Their simultaneous determination is therefore essential for a complete description of the vector current and provides the electromagnetic input required for weak single-pion production.
providing comprehensive coverage of the resonance and transition regions.

                        \begin{table}
                          \centering
                          \caption{Summary of the data sets used in the analysis to determine the proton and neutron form factors. The data span a broad kinematic range and include multiple channels from electron- and photon-scattering measurements. The photon-scattering data are compiled from several experiments performed with different beam energies.}
                          \label{clas_data}
                          \renewcommand{\arraystretch}{1.3}
                          \begin{ruledtabular}
                            \begin{tabular}{lccl}
                              Channel & $E_e~\text{GeV}$&$Q^2~(\text{GeV}/c)^2$& $W~\text{GeV}$ 
                              \\ [0.1ex]
                              \hline
                              $e p \rightarrow e p \pi^{0}$ & 1.046  & 0.16 - 0.32  & 1.1 - 1.34 \\
                              $e p \rightarrow e n \pi^{+}$ & 1.046  & 0.16 - 0.32  & 1.1 - 1.34\\
                              $e p \rightarrow e n \pi^{+}$ & 1.515  & 0.30 - 0.60 & 1.11 - 1.57\\
                              $e p \rightarrow e p \pi^{0}$ & 1.645  & 0.40 - 0.90  & 1.1 - 1.68\\
                              $e p \rightarrow e p \pi^{0}$ & 2.445  & 0.65 - 1.80  & 1.1 - 1.68\\
                              $e p \rightarrow e n \pi^{+}$ & 5.499  & 1.80 - 4.00 & 1.62 - 2.01\\
                              $e p \rightarrow e n \pi^{+}$ & 5.754  & 1.72 - 4.16 & 1.15 - 1.67\\
                              $e p \rightarrow e p \pi^{0}$ & 5.754  & 3.00 - 6.00 & 1.11 - 1.39\\
                              $e n \rightarrow e p \pi^{-}$ & 2.039  & 0.4 - 1.0 & 1.1 - 1.8\\
                              $\gamma n \rightarrow p \pi^{-}$ & -  & 0.00 & 1.1 - 2.01\\    
                              $\gamma p \rightarrow  n \pi^{+}$ & -  & 0.00 & 1.1 - 2.01\\                     \end{tabular}
                          \end{ruledtabular}
                        \end{table}
        \subsection {Pion-Nucleon and Neutrino-Nucleon Single Pion Productions}
        
    Neutrino weak interactions involve both vector and axial-vector currents. Assuming that the vector current is accurately constrained through electromagnetic data included in the joint analysis, neutrino scattering data can be used to determine the axial-vector form factors. The SPP channels induced by neutrino and antineutrino beams proceed either via charged-current (CC) interactions,
             
    \begin{eqnarray}
    \nu_{l} + p \rightarrow l^- p ~\pi^{+}~~&,&~~~~~~~~~ \bar{\nu}_{l} + n \rightarrow  l^+ n ~\pi^{-} \nonumber\\
    \nu_l  + n \rightarrow l^- n ~\pi^{+}~~&,&~~~~~~~~~ \bar{\nu}_l + p \rightarrow  l^+ p ~\pi^{-} \nonumber\\
    \nu_l + n \rightarrow l^- p ~\pi^{0}~~&,&~~~~~~~~~ \bar{\nu}_l + p \rightarrow  l^+ n ~\pi^{0},
    \end{eqnarray}
    or via neutral-current (NC) interactions,
    \begin{eqnarray}
    \nu_{l} + p \rightarrow \nu_{l} p ~\pi^{0}~~&,&~~~~~~~~~ \bar{\nu}_{l} + p \rightarrow  \bar{\nu}_{l} p ~\pi^{0} \nonumber\\
    \nu_{l} + p \rightarrow \nu_{l} n ~\pi^{+}~~&,&~~~~~~~~~ \bar{\nu}_{l} + p \rightarrow  \bar{\nu}_{l} n ~\pi^{+} \nonumber\\
    \nu_{l} + n \rightarrow \nu_{l} n ~\pi^{0}~~&,&~~~~~~~~~ \bar{\nu}_{l} + n \rightarrow  \bar{\nu}_{l} n ~\pi^{0} \nonumber\\
    \nu_{l} + n \rightarrow \nu_{l} p ~\pi^{-}~~&,&~~~~~~~~~ \bar{\nu}_{l} + n \rightarrow  \bar{\nu}_{l} p ~\pi^{-}.
    \end{eqnarray}

     Experimental knowledge of the neutrino--nucleon cross section for single-pion production in the GeV neutrino-energy range, which is crucial for precision predictions, remains limited. Most of the existing data originate from low-statistics bubble chamber experiments performed on hydrogen or deuterium targets using muon (anti)neutrino beams.
    
    Unlike electron-scattering experiments, where the full kinematics are typically measured, neutrino data are generally reported either as integrated cross sections as a function of neutrino energy or as one-dimensional differential distributions, most commonly $d\sigma/dQ^2$. The only measurements on hydrogen targets come from CC neutrino and antineutrino channels recorded by the BEBC experiment, which employed broad neutrino and antineutrino beams with an average energy of approximately $E_{\nu} \sim 20~\text{GeV}$. Consequently, no measurements on hydrogen exist in the few-GeV neutrino-energy range relevant for current and future accelerator-based neutrino experiments.
    
    Bubble chamber measurements in this energy region were instead performed by the ANL and BNL experiments, with an average neutrino energy of approximately $E_{\nu} \sim 1~\text{GeV}$, using deuterium targets. However, nuclear effects in deuterium modify the cross section at the level of approximately $10$--$15\%$~\cite{CLAS:2022kta}. Since data in the few-GeV region are essential for this study, and because the model parametrisation developed here is designed to describe both non-perturbative and perturbative regimes, I include both the ANL and BEBC data sets in the analysis. To account for nuclear effects in the deuterium measurements, the ANL data are assigned an additional $15\%$ uncertainty.

    As neutrino data provide only limited constraints on the axial-vector form factors, I adopt a parametrisation analogous to that used for the vector form factors in the quark-hadron transition region. In particular, the parameters introduced for $L_i(Q^2)$ in Eq.~\ref{log} are taken to be identical for both the vector and axial-vector form factors. $F_i^A(0)$, The other factors in the axial-vector form factors in Eq. \ref{FFCg}, can be constrained using pion-nucleon scattering data on hydrogen.
    
    The partially conserved axial current (PCAC) hypothesis~\cite{Adler} relates the divergence of the hadronic axial current near $Q^2=0$ to the pion field operator. Consequently, it provides a connection between weak SPP and elastic pion-nucleon scattering, leading to the following relation:
        
                \begin{eqnarray} \label{pcac}
                    \frac{d\sigma}{dWdQ^2}(\nu_{\mu} p \rightarrow \mu^- p \pi^+) \Bigm\lvert_{Q^2\to 0} &=& \mathcal{A}\sigma(\pi^+ p \rightarrow \pi^+ p)\nonumber\\
                    \frac{d\sigma}{dWdQ^2}(\bar{\nu}_{\mu} p \rightarrow \mu^+ p \pi^-) \Bigm\lvert_{Q^2\to 0} &=& \mathcal{A}\sigma(\pi^- p \rightarrow \pi^- p)\nonumber\\
                \end{eqnarray}   
        Where 
        \begin{eqnarray}
        \mathcal{A} &=& G_F^2 \cos^2\theta_C \left(\frac{f_\pi}{\pi}\right)^2 \frac{|\mathbf{q}|}{|\mathbf{k}|^2} \frac{E_l}{E}\Bigg[ \left|\frac{k_0}{|\mathbf{k}| }- \frac {m_l^2|\mathbf{k}|}{2k_{20}}\frac{1}{k^2 + m_{\pi}^2}\right |^2 \nonumber\\
            &~& + \frac{m_l^2\theta^2}{4}\left(\frac{|\mathbf{k}|}{k^2 + m_{\pi}^2}\right)^2\Bigg]
        \end{eqnarray}
            
        Here $\sigma(\pi^{\pm} p \rightarrow \pi^{\pm} p)$ is the pion-nucleon cross-section with the initial pion off the mass shell i.e. the Eq. \ref{pcac} is a relation between observable quantities if the difference between $Q^2$ and $m_{\pi}^2$ can be ignored.

         \section{RESULTS AND COMPARISON WITH
        EXPERIMENTAL DATA} 
    
        The complete model contains 95 free parameters relevant for form-factors defined in 6 resonances and non-resonant as introduced in section \ref{FF}, including $F_i^{(V,A)}(0)$, $C_i^{(V,A)}(0)$, and parameters $a_{ik}^(V,A)$ after applying the linear superconvergence relations and parameters $K_{i_j},~j=0,2,4$ defined in the lagrangian renormalisation $L_i(k^2)$. I also introduced phases between resonant and non-resonant amplitudes that is relevant for the interfrence terms. These parameters are determined through a global fit to the experimental data described in the previous section, using the same methodology as in my previous works \cite{MK2,MK3}. The resulting best-fit solution yields a reduced $\chi^2 \approx 1$, indicating excellent overall agreement with the data.
    
    In this section, I present representative results of the fit. The selected cross sections are taken from multiple reaction channels and experimental data sets, spanning a broad range of $Q^2$ and $W$ values. This demonstrates the validity of the MK model predictions over the wide kinematic region relevant to accelerator-based neutrino experiments. All results are shown together with their corresponding $1\sigma$ uncertainty bands. To obtain a reliable estimate of the systematic uncertainties, correlations among the fitted parameters are fully taken into account, thereby avoiding both double counting and artificial cancellations of systematic effects.\footnote{The model has been implemented in the NEUT neutrino interaction event generator~\cite{Hayato:2021heg}. The corresponding best-fit parameters and covariance matrix are publicly available in ROOT format.}

         \begin{figure*}
                  \centering
                  {\begin{minipage}{1.06\textwidth}
                      \includegraphics[width=\textwidth]{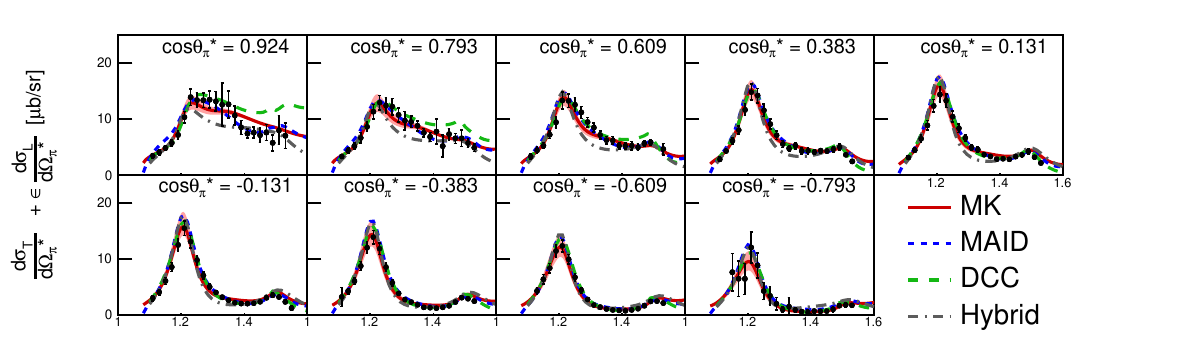}
                          \vspace*{-6.ex}
                    \end{minipage}
                    \begin{minipage}{1.06\textwidth}
                      \includegraphics[width=\textwidth]{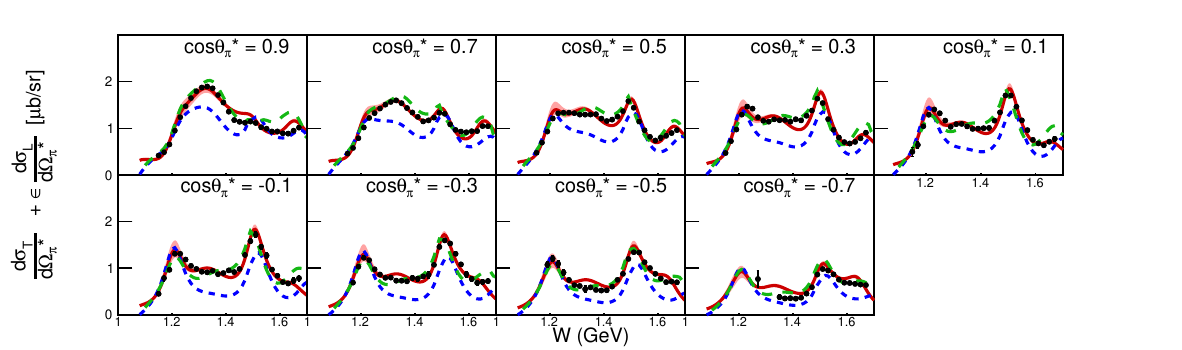}
                 \end{minipage}
                  }
                 \caption{Differential cross section as a function of the invariant hadronic mass for the $e p \rightarrow e n \pi^{+}$ channel. The top panels correspond to $E = 1.515~\text{GeV}$ and $Q^2 = 0.4~(\text{GeV}/c)^2$, while the bottom panels correspond to $E = 5.754~\text{GeV}$ and $Q^2 = 2.05~(\text{GeV}/c)^2$. The MK model prediction is shown by the solid red line, with the $68\%$ confidence interval indicated by the shaded band. The MAID, DCC, and Hybrid model predictions are shown by the dashed blue, long-dashed green, and dash-dotted grey lines, respectively. The data and the definition of the cross section are taken from Refs.~\cite{pi+, pi+_high}.}
        \label{npip}
                \end{figure*}
         \begin{figure*}
                  \centering             \includegraphics[width=1.1\textwidth]{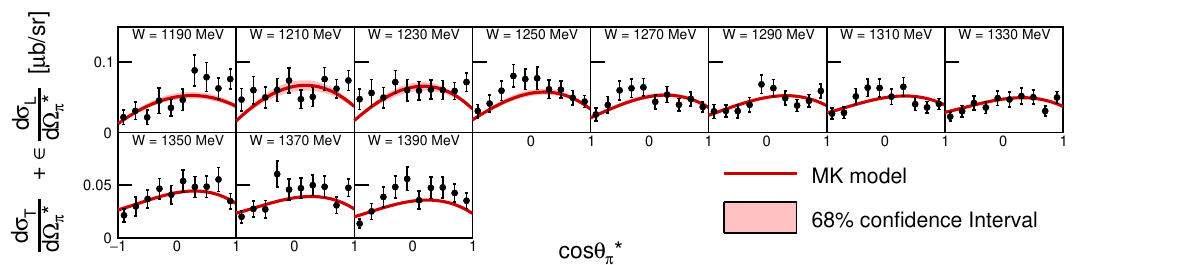}
                
                  \caption{Differential cross section as a function of the pion polar angle in the isobaric frame for the $e p \rightarrow e p \pi^{0}$ channel at $E_e = 5.754~\text{GeV}$ and $Q^2 = 6.0~(\text{GeV}/c)^2$, shown for various invariant hadronic masses. The MK model prediction is shown by the solid red line, with the $68\%$ confidence interval indicated by the shaded band. The data and the definition of the cross section are taken from Ref.~\cite{pi0_high}.}
         \label{pi0}
                \end{figure*}    
          Figures~\ref{npip} and \ref{pi0} show differential cross-section measurements in electron--proton scattering at low, medium, and high $Q^2$, together with the MK model predictions. Several other models describe electron--proton interactions, primarily in the low-$Q^2$ region; however, to my knowledge, no model predictions are available for $Q^2 = 6.0~(\text{GeV}/c)^2$, as shown in Fig.~\ref{pi0}. In Fig.~\ref{npip}, I also present predictions from the MAID~\cite{MAID2007}, Dynamical Coupled-Channels (DCC)~\cite{Nakamura:2015rta}, and Hybrid~\cite{gonzalez-jimenez2019} models. A summary of these models can be found in my previous work~\cite{MK3}. It is worth noting that the MAID and DCC models determine the parameters of their electromagnetic SPP models by fitting to electron- and photon-scattering data.        
        
\begin{figure*}
                  \centering
                  
                      \includegraphics[width=\textwidth]{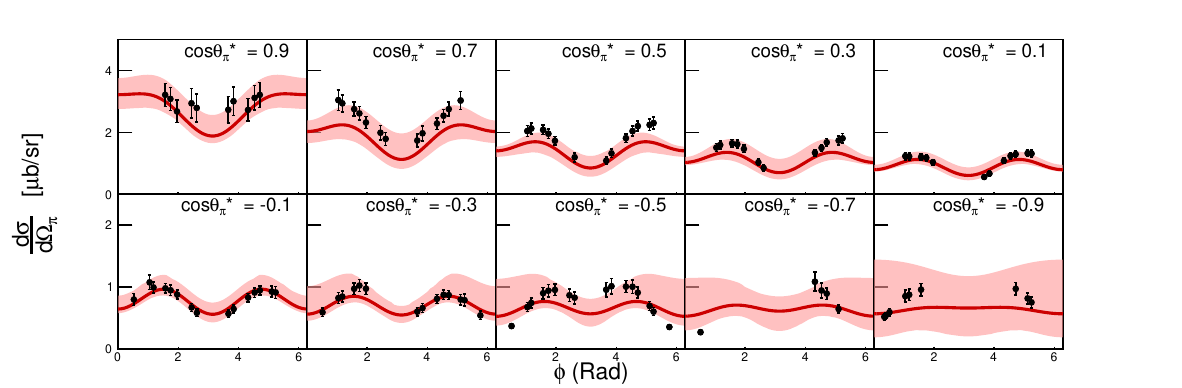}
                
                  \vspace*{-2ex}
                   \caption{Differential cross section for the $e n \rightarrow e p \pi^{-}$ channel as a function of the azimuthal pion angle in the isobaric frame at $E_e = 2.039~\text{GeV}$, $W = 1.4875~\text{GeV}$, and $Q^2 = 0.7~(\text{GeV}/c)^2$, shown for various pion polar angles in the isobaric frame. The MK model prediction is shown by the solid red line, with the $68\%$ confidence interval indicated by the shaded band. The data and the definition of the cross section are taken from Ref.~\cite{CLAS:2022kta}.}\label{pim}
                \end{figure*}

        Figure~\ref{pim} shows differential cross-section measurements in electron--neutron interactions in the second resonance region, where the excited neutron states have form factors that differ from those of the proton. Understanding these form factors is essential for determining the vector form factors relevant to weak interactions.
    
\begin{figure*}
                  \centering
                 {\begin{minipage}{0.49\textwidth}
        \includegraphics[width=\textwidth]{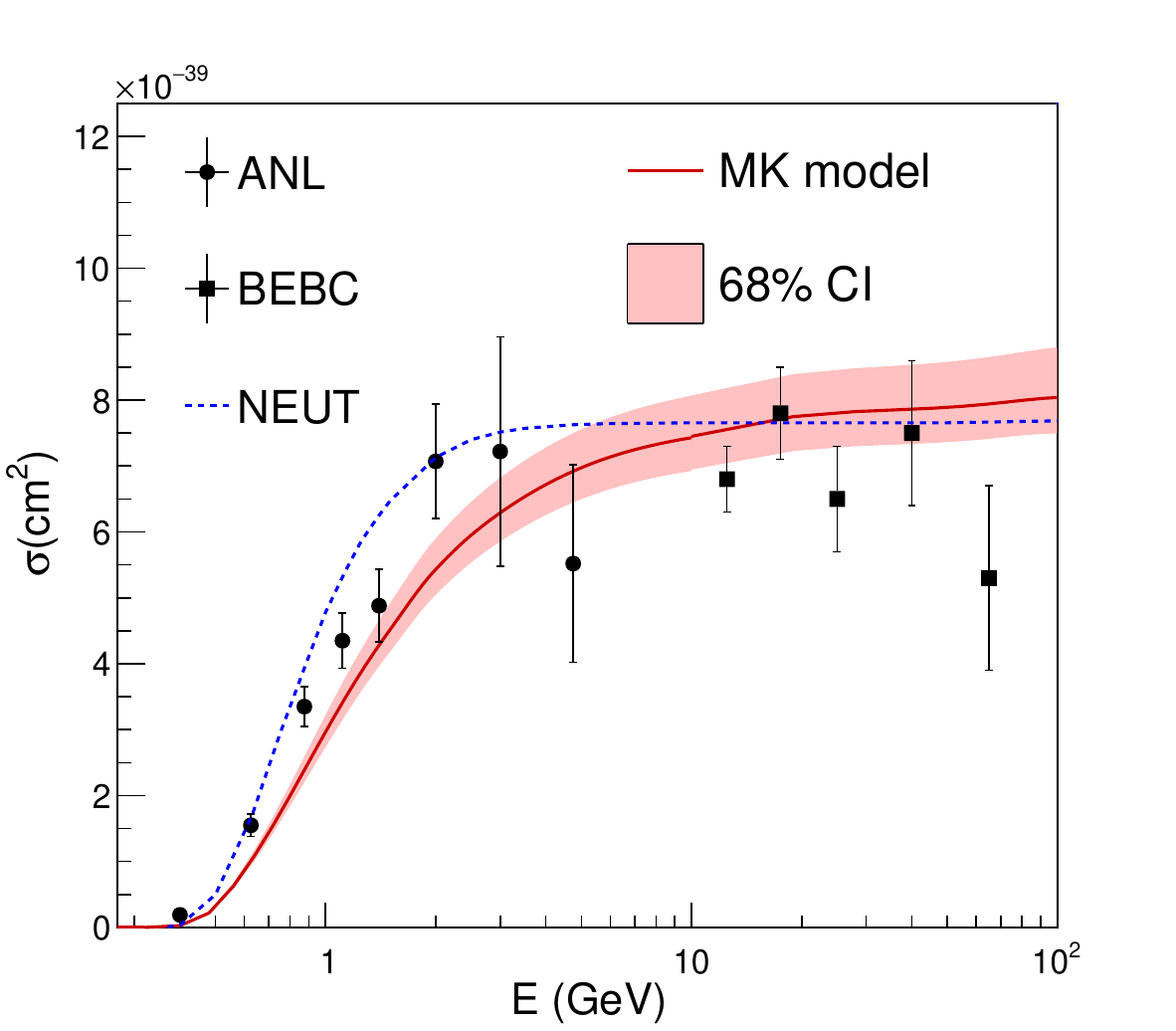}
                    \end{minipage}
                    \begin{minipage}{0.49\textwidth}
                      \includegraphics[width=\textwidth]{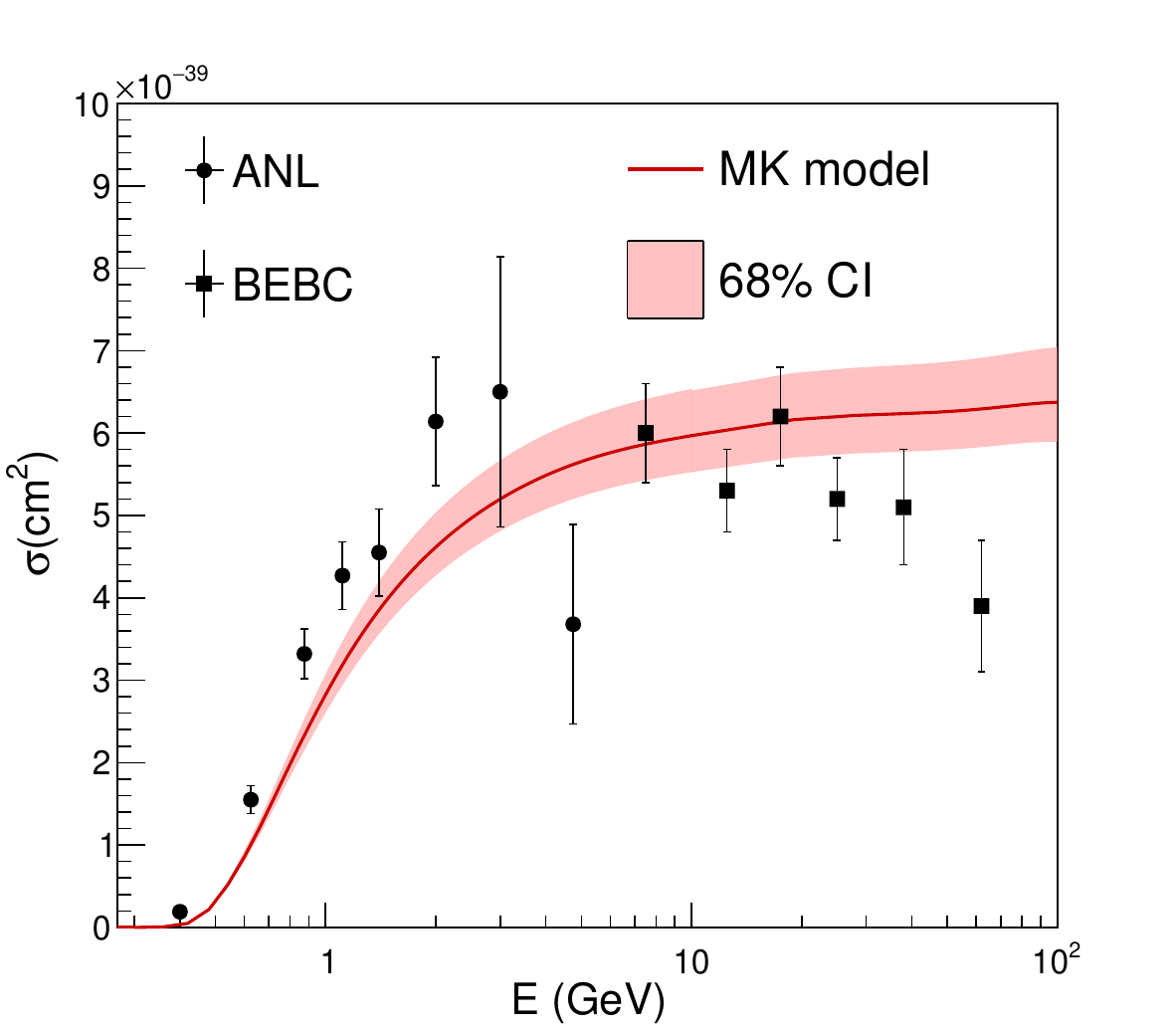}
                 \end{minipage}
                  }      
                
                  \caption{Integrated cross section for the $\nu p \rightarrow \mu^- p \pi^{+}$ channel as a function of neutrino energy for $W < 2~\text{GeV}$ (left) and $W < 1.4~\text{GeV}$ (right). The MK model prediction is shown by the solid red line, with the $68\%$ confidence interval indicated by the shaded band. The prediction from the NEUT event generator is shown by the dashed blue line. The ANL and BEBC data are taken from Refs.~\cite{Radecky:1981fn} and \cite{Aachen-Bonn-CERN-Munich-Oxford:1980iaz}.} \label{ppip_E}
                \end{figure*}

\begin{figure*}
                  \centering
                 {\begin{minipage}{0.49\textwidth}
                      \includegraphics[width=\textwidth]{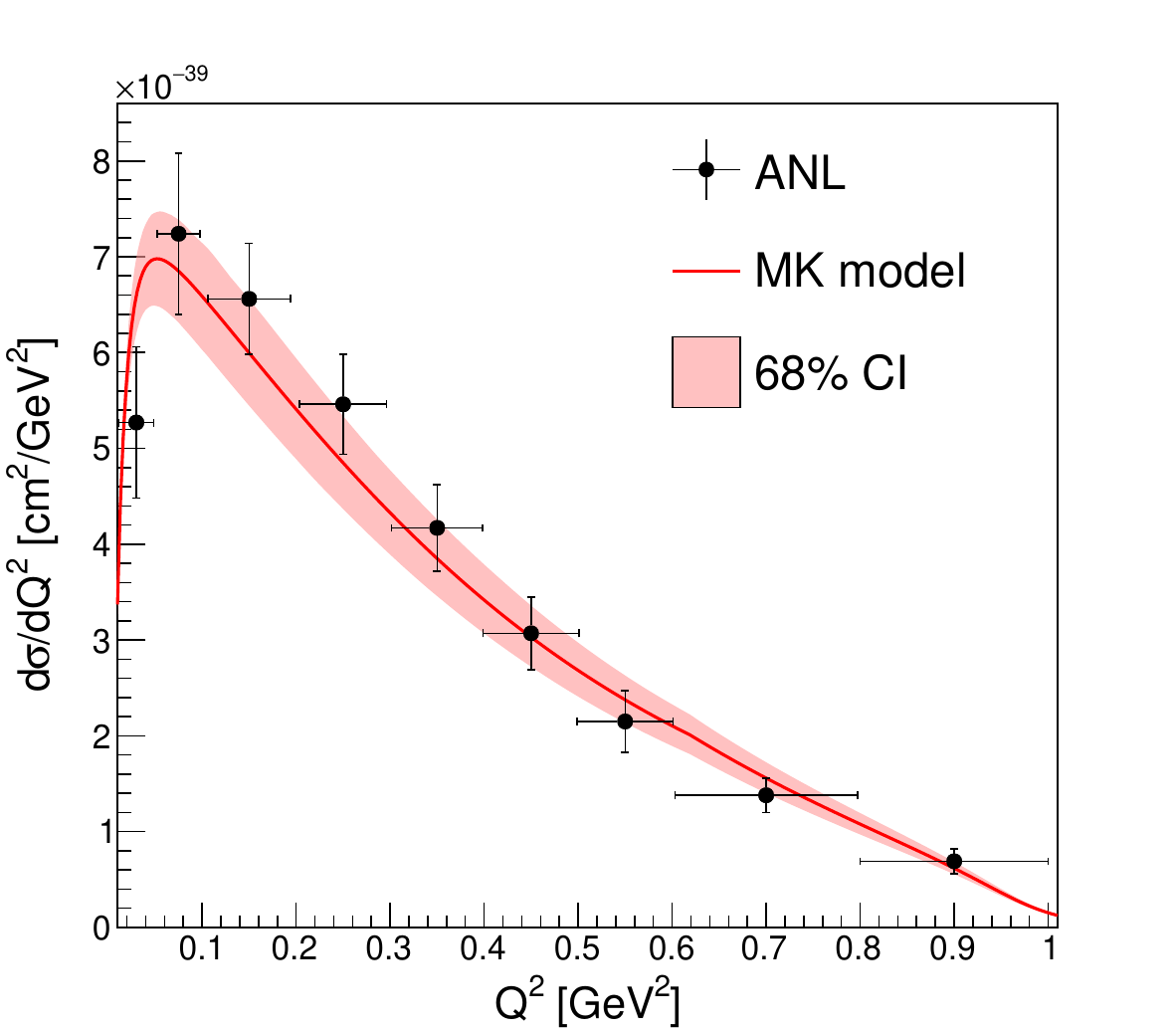}
                    \end{minipage}
                    \begin{minipage}{0.49\textwidth}
                      \includegraphics[width=\textwidth]{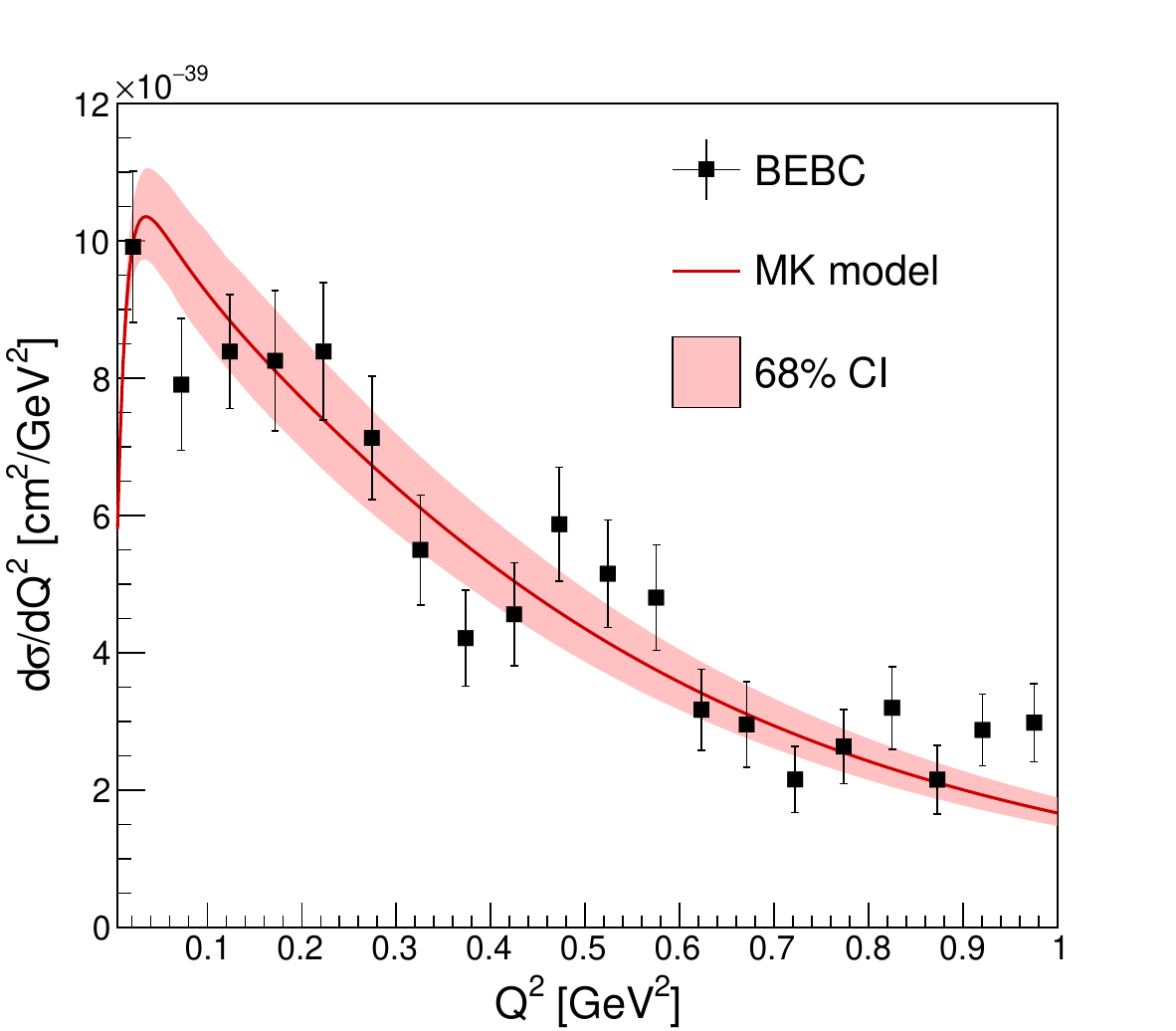}
                 \end{minipage}
                  }      
                
                  \caption{Differential cross section $d\sigma/dQ^2$ for the $\nu p \rightarrow \mu^- p \pi^{+}$ channel as a function of $Q^2$ for $W < 1.4~\text{GeV}$. The MK model prediction (solid red line) is compared with ANL (left) and BEBC (right) data. The prediction from the NEUT event generator is shown by the dashed blue line. The ANL and BEBC data are taken from Refs.~\cite{Radecky:1981fn} and \cite{WA21:1989gku}.}
        \label{ppip_Q}
                \end{figure*}

\begin{figure*}
                  \centering
                 {\begin{minipage}{0.49\textwidth}
                      \includegraphics[width=\textwidth]{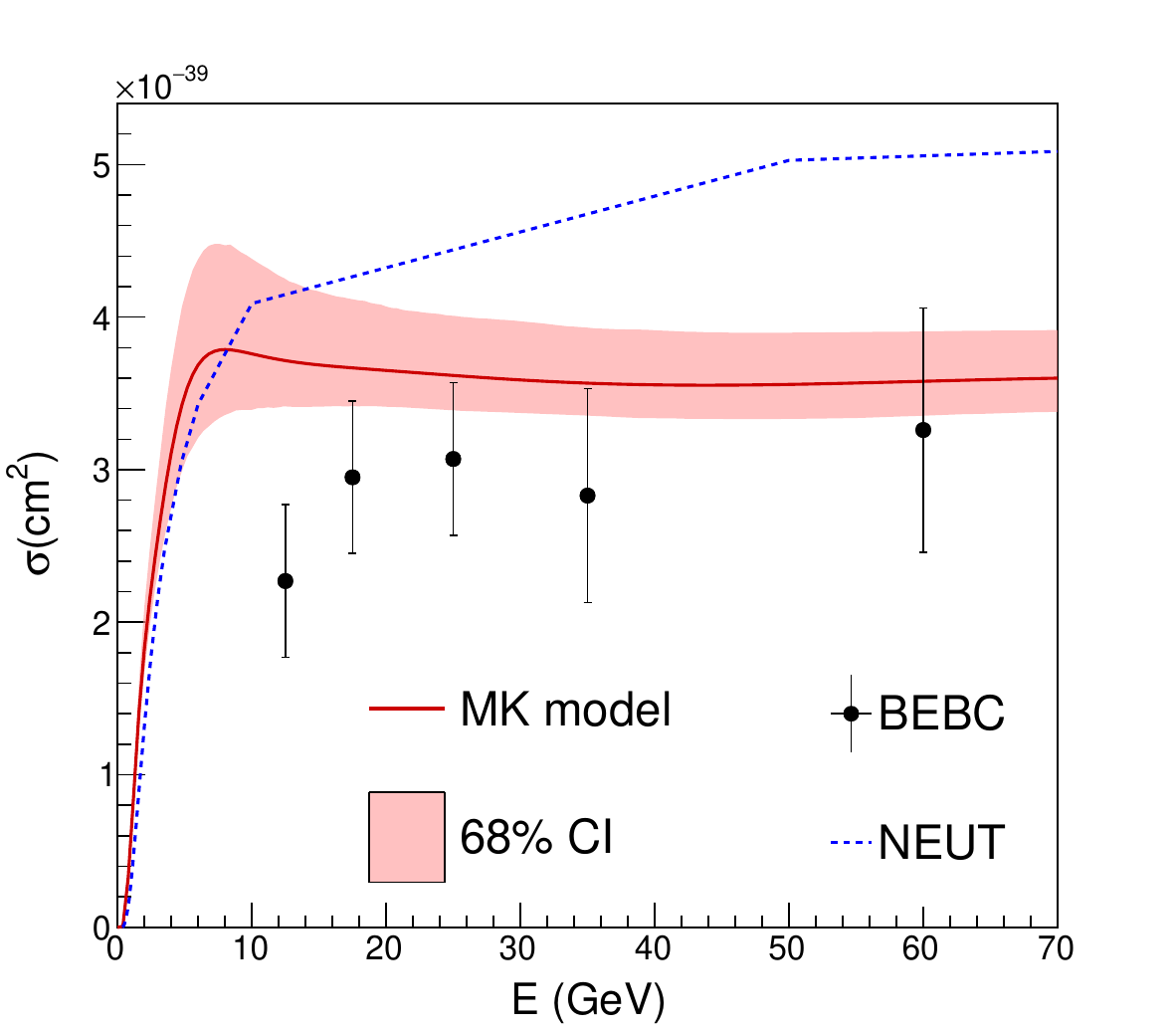}
                    \end{minipage}
                    \begin{minipage}{0.49\textwidth}
                      \includegraphics[width=\textwidth]{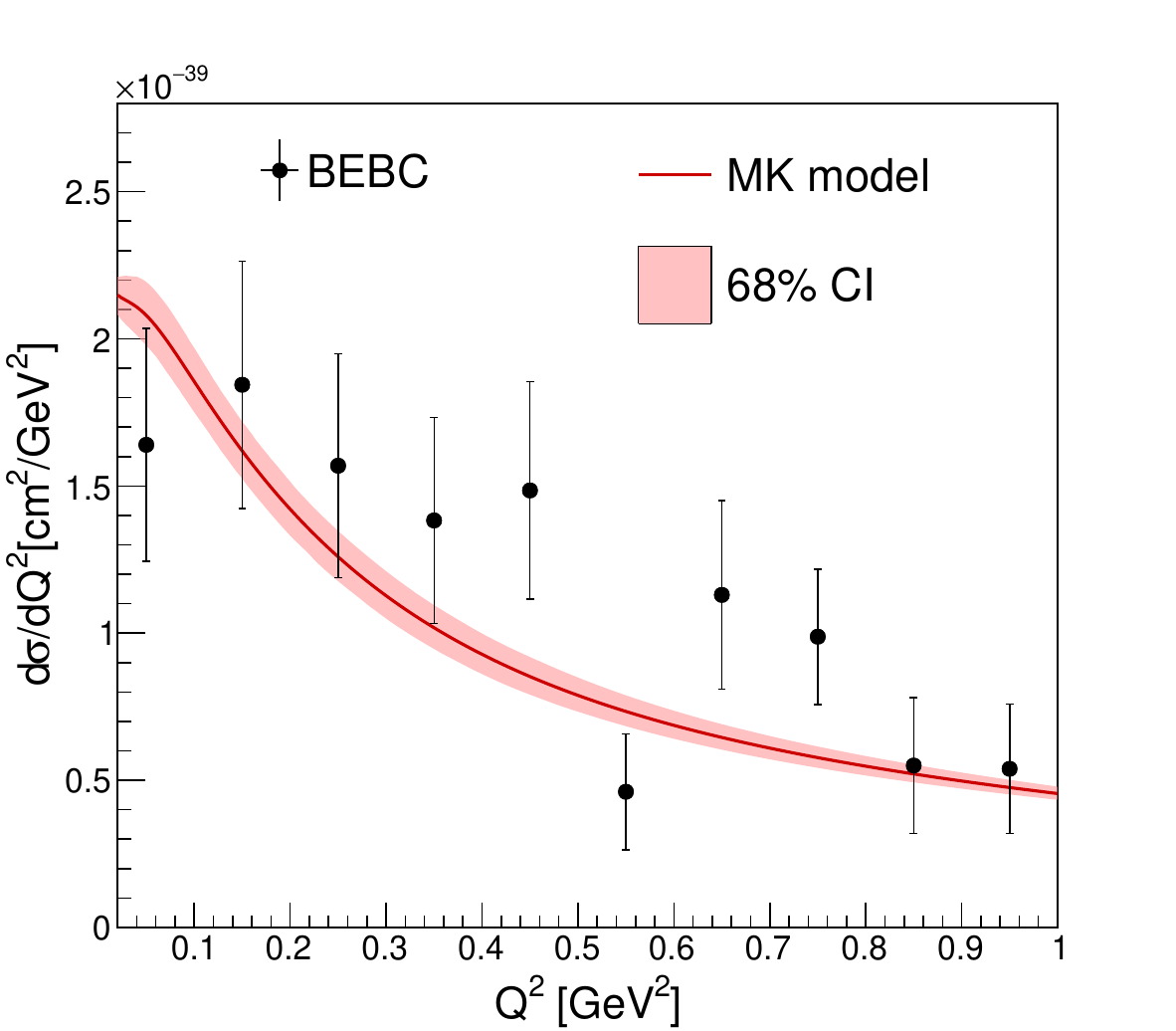}
                 \end{minipage}
                  }      
                  \caption{Comparison of the MK model prediction with BEBC data for the $\bar{\nu} p \rightarrow \mu^{+} p \pi^{-}$ channel. The left panel shows the integrated cross section as a function of neutrino energy for $W < 2~\text{GeV}$, while the right panel shows the differential cross section $d\sigma/dQ^2$ for $W < 1.4~\text{GeV}$. The MK model prediction is shown by the solid red line, with the $68\%$ confidence interval indicated by the shaded band. The prediction from the NEUT event generator is shown by the dashed blue line. The BEBC data are taken from Refs.~\cite{Aachen-Bonn-CERN-Munich-Oxford:1980iaz} and \cite{WA21:1989gku}.}
         \label{PPanu}
                \end{figure*}

\begin{figure*}
                  \centering
                  {\begin{minipage}{0.49\textwidth}
                      \includegraphics[width=\textwidth]{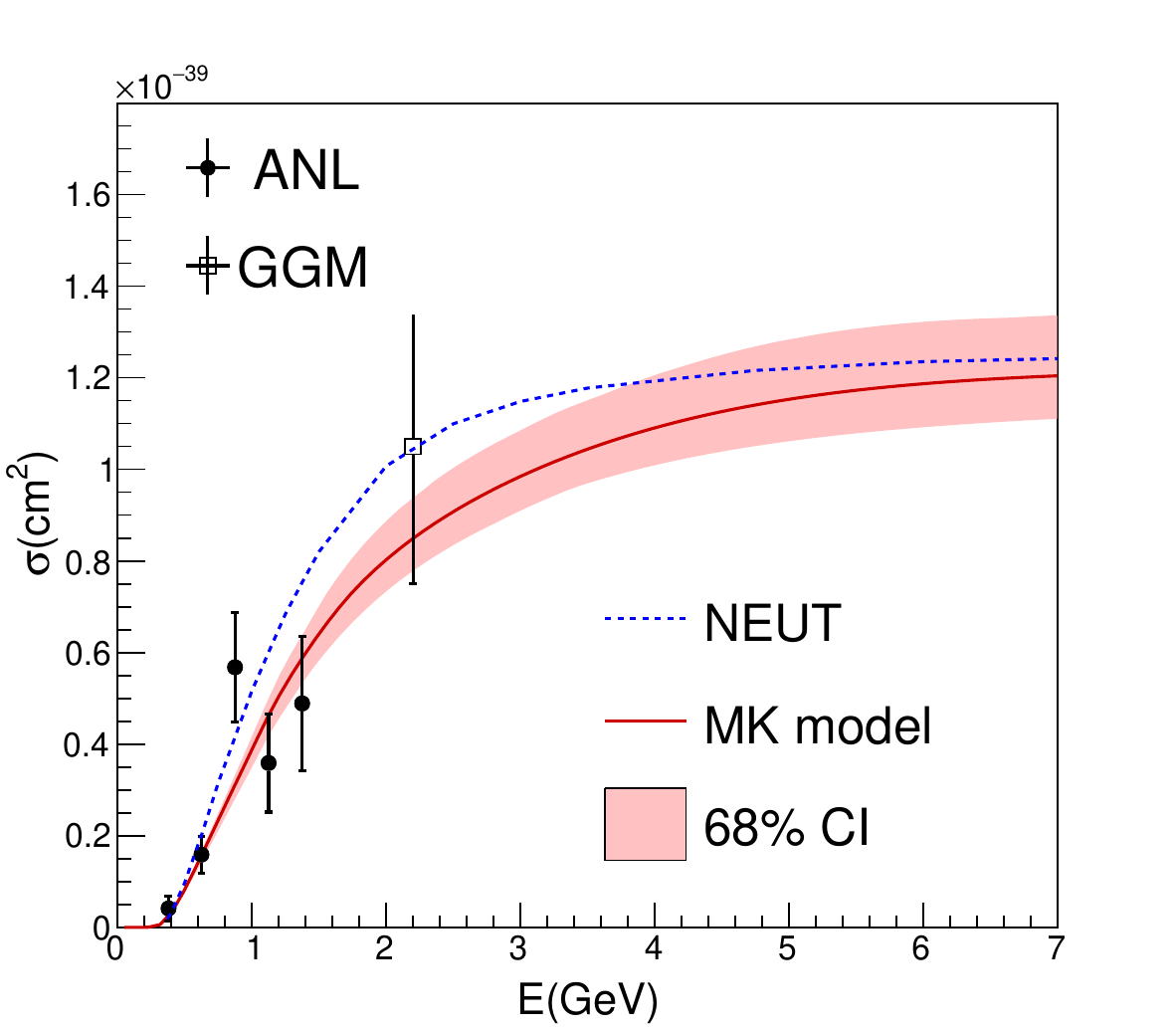}
                    \end{minipage}
                    \begin{minipage}{0.49\textwidth}
                      \includegraphics[width=\textwidth]{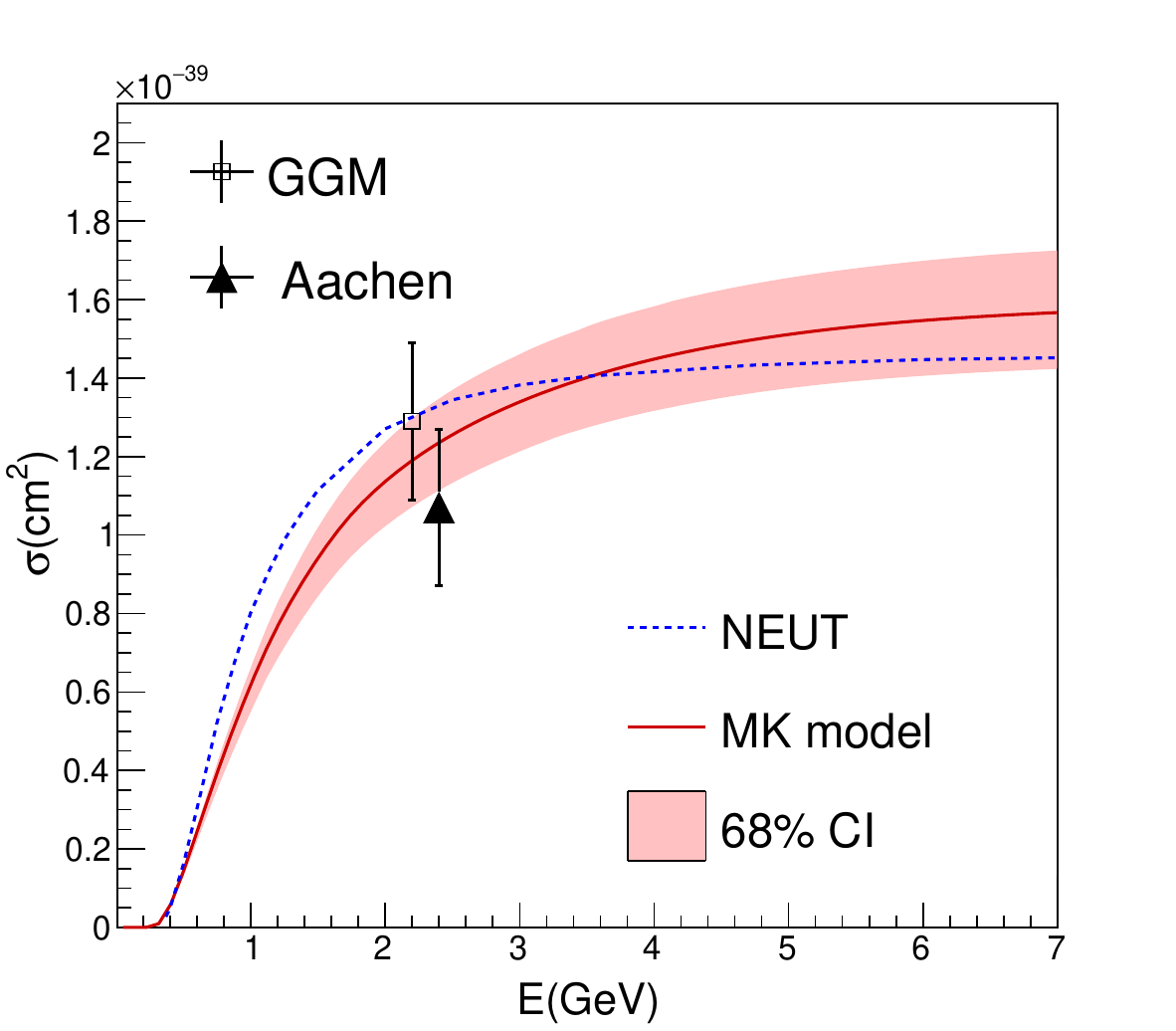}
                 \end{minipage}
                  }
                 \caption{Integrated cross sections in the $W < 2~\text{GeV}$ region for the $\nu n \rightarrow \nu p \pi^{-}$ (left) and $\nu p \rightarrow \nu p \pi^{0}$ (right) channels as a function of neutrino energy. The MK model prediction is shown by the solid red line, with the $68\%$ confidence interval indicated by the shaded band. The prediction from the NEUT event generator is shown by the dashed blue line. The data are taken from Refs.~\cite{Derrick:1980nr, GargamelleNeutrinoPropane:1977hya, Faissner:1983ng}.}
        
        \label{NC}
        \end{figure*}

        The neutrino data described in the previous section are shown in Figs.~\ref{ppip_E}-\ref{NC} for the ANL and BEBC experiments. To my knowledge, the MK model provides the first complete SPP framework capable of simultaneously describing both data sets. This is achieved through a unified modelling approach combined with a global analysis that consistently treats the quark-hadron transition region. 
    
    This capability is particularly important because the BEBC experiment provides the only available measurements of differential cross sections for both neutrino and antineutrino beams. Achieving simultaneous control over neutrino and antineutrino channels is essential for future precision measurements of CP violation in accelerator-based neutrino experiments.
    
    For comparison, integrated cross-section predictions from the NEUT~5.3.6 neutrino event generator~\cite{Hayato:2021heg} are also shown. NEUT is widely used to simulate neutrino interactions in oscillation experiments such as T2K in Japan. The SPP model implemented in NEUT is based on the Rein--Sehgal model developed in 1981~\cite{RS}.

    Figure~\ref{ppip_E} shows the integrated cross section for the $\nu p \rightarrow \mu^- p \pi^{+}$ channel measured by the ANL and BEBC experiments with invariant-mass selections of $W < 2.0~\text{GeV}$ and $W < 1.4~\text{GeV}$. This channel contains only resonances with isospin $3/2$; therefore, in the right panel the dominant contribution arises from the $\Delta$ resonance.

        Differential cross-section measurements $d\sigma/dQ^2$ with (anti)neutrino beams are essential for extracting the $Q^2$ dependence of form factors in weak interactions, particularly the axial form factors, which cannot be constrained by electron-scattering data. Figure~\ref{ppip_Q} shows the comparison between the data and the MK model predictions for $d\sigma/dQ^2$ in the $\nu p \rightarrow \mu^- p \pi^{+}$ channel measured by the ANL (right) and the BEBC (left) experiments. 
        
        Figure~\ref{PPanu} shows the corresponding BEBC measurements for the $\bar{\nu} p \rightarrow \mu^{+} p \pi^{-}$ channel. this reaction receives contributions from both isospin-$1/2$ and isospin-$3/2$ resonances, allowing the parameters of the axial current associated with isospin-$1/2$ resonances to be constrained.
        
        There are extremely limited data available for some neutral-current channels. Although these data are not included in the joint analysis, they nevertheless show reasonable agreement with the MK model predictions, as illustrated in Fig.~\ref{NC}.
    
    \section{Conclusion}
                 
    The single-pion production (SPP) model presented in this work is designed to improve our understanding of neutrino interactions and to support the interpretation of neutrino measurements. The main features of the MK model are summarised below:
    
    \begin{itemize}
    
    \item {\bf Comprehensive Coverage:} The model includes all SPP channels and is applicable across the full kinematic range relevant to the energy scales of current and future accelerator-based neutrino experiments.
    
    \item {\bf Neutral-Current Weak Interactions:} By employing a unified description of electromagnetic and weak interactions together with isospin symmetry, the model provides reliable predictions for neutral-current (NC) neutrino interactions, even in the absence of extensive experimental data. This capability is particularly important for water Cherenkov detectors such as T2K and Hyper-K, where NC $\pi^0$ production constitutes a significant background.
    
    \item {\bf Description of the Low-$Q^2$ Region:} The model addresses the low-$Q^2$ region, where current discrepancies with data are often observed. This is achieved through the use of photon scattering data and the Conserved Vector Current (CVC) hypothesis for the vector current, together with pion scattering data and the Partially Conserved Axial Current (PCAC) hypothesis for the axial current.
    
    \item {\bf Behaviour at High $Q^2$:} The neutrino--nucleon resonance vertices are parametrised using phenomenological form factors that satisfy analyticity and unitarity constraints. These form factors are constructed to be asymptotically consistent with QCD expectations, ensuring the validity of the model at high $Q^2$.
    
    \item {\bf Consistent Description of Neutrino and Antineutrino Channels:} The model provides a simultaneous description of both neutrino and antineutrino interactions. Achieving comparable accuracy for both channels is essential for future precision measurements of CP violation in long-baseline neutrino experiments.

    \item {\bf Integration with Event Generators:} The framework is designed for straightforward implementation within neutrino interaction event generators and nuclear theory calculations used in modern neutrino experiments.
    
    \item {\bf Control of Systematic Uncertainties:} The analysis incorporates careful treatment of systematic uncertainties arising from experimental errors, model assumptions, parametrisation choices, and extrapolation to higher energies. Such control is essential for achieving the precision required in modern neutrino measurements.
    
    \end{itemize}
                
    By addressing these key aspects, the MK model provides a robust and comprehensive framework for the study of neutrino and antineutrino interactions. This work contributes to improving the reliability of neutrino interaction modelling and supports the precision measurements required for future discoveries, including the search for leptonic CP violation.

\begin{acknowledgments}
  I am extremely grateful to Dave Wark, Deborah Harris, and Kevin McFarland for their constant encouragement, support, and invaluable advice throughout my academic career. I also thank Luis Alvarez Ruso, Clarence Wret, Philip Rodrigues, Lukas Koch, Stephen Dolan, Laura Munteanu, Dan Cherdack, and Callum Wilkinson for many fruitful discussions. I am grateful to my colleagues in the neutrino physics community at Imperial College London, the University of York, the University of Oxford, and Fermilab for their support and for providing a stimulating research environment.

This work was supported by the Royal Commission for the Exhibition of 1851 through a Research Fellowship and by the Royal Society through a University Research Fellowship (grant no. URF\textbackslash R1\textbackslash 251471).
\end{acknowledgments}                  
        \appendix
            \section{Helicity amplitudes for the resonant interactions}\label{appA}
            The helicity amplitudes of the vector and axial-vector currents are given in Table~\ref{res_HV} and Table~\ref{res_HAA}, where $f^{V,A}(R)$ is the production amplitudes and $D(R)$ is the decay amplitude:
        
            \begin{align}
                        \mathcal{D}^j(R)=\langle N\pi,\lambda_2|R \lambda_R \rangle &= \sigma^D C_{N\pi}^{j} \sqrt{\chi_E} \kappa C_{N\pi}^{I} f_{BW} 
                        \end{align}
                        where $f_{BW}(R)$ is the Breit-Wigner amplitude:
                        \begin{eqnarray}
                        f_{BW}(R) = \sqrt{\frac{\Gamma_R}{2\pi} }\left( \frac{1}{W- M_R + i\Gamma_R/2} \right )
                        \end{eqnarray}
                        where \begin{eqnarray}
                        \Gamma_R = \Gamma_0 (|\mathbf{q}(W)|/|\mathbf{q}(M_R)|)^{2l+1},
                        \end{eqnarray}
                        and 
                        \begin{equation}
                        \kappa= \left( 2\pi^2 \frac{W^2}{M^2} ~.~\frac{2}{2j+1} ~ \frac{1}{|\mathbf{q}|} \right)^{\frac{1}{2}},
                        \end{equation}
                        $M_R$, $\Gamma_0$, $\chi_E$ are resonance's mass, width and the branching ratio \cite{ParticleDataGroup:2024cfk} and $C^I_{N\pi}$ are the isospin Clebsch-Gordan coefficients given in Table~\ref{RSisospin}.

     \begin{table}
                        \centering
                        \caption{Isospin coefficients} \label{RSisospin}
                        \renewcommand{\arraystretch}{1.5}
                        \begin{ruledtabular}
                         \begin{tabular}{|ccc|}
                         {Channels} & $C^{3/2}_{N\pi}$& $C^{1/2}_{N\pi}$   \\ [0.7ex]
                         \hline
                        $e p \rightarrow e p \pi^{0}$    &  $\sqrt{\frac{2}{3}}$  &  $-\sqrt{\frac{1}{3}}$\\
                        $e p \rightarrow e n \pi^{+}$   &$-\sqrt{\frac{1}{3}}$  &$-\sqrt{\frac{2}{3}}$\\
                        \end{tabular}
                        \end{ruledtabular}
                        \end{table}
                        The explicit form of the $d^j_{\lambda,\mu}(\theta)$ functions for $j=l+\frac{1}{2}$ are:
                        \begin{align}
                        d^j_{\frac{1}{2} \frac{1}{2}}~&= (l+1)^{-1} \cos\frac{\theta}{2} (P'_{l+1} - P'_l)\nonumber\\
                        d^j_{-\frac{1}{2} \frac{1}{2}}&= (l+1)^{-1} \sin\frac{\theta}{2} (P'_{l+1} + P'_l)\nonumber\\
                        d^j_{\frac{1}{2} \frac{3}{2}}~&= (l+1)^{-1} \sin\frac{\theta}{2} (\sqrt{\frac{l}{l+2}}P'_{l+1} + \sqrt{\frac{l+2}{l}} P'_l)\nonumber\\
                        d^j_{-\frac{1}{2} \frac{3}{2}}&= (l+1)^{-1} \cos\frac{\theta}{2} (-\sqrt{\frac{l}{l+2}}P'_{l+1} + \sqrt{\frac{l+2}{l}} P'_l)\nonumber                   \end{align}
                        where $P_l$ are Legendre polynomials and $P'_l= dP_l/d\cos\theta$.
                        \begin{table*}
                        \centering
                        \caption{Vector helicity amplitudes of resonant interaction. }
                        \label{res_HV}
                        \renewcommand{\arraystretch}{1.3}
                        \begin{ruledtabular}
                         \begin{tabular}{c|c|c|c}
                          $\lambda_2$& $\lambda_1$ &$\tilde{F}_{\lambda_2 \lambda_1}^{e_{L}}(\theta, \phi)$& $\tilde{F}_{\lambda_2 \lambda_1}^{e_{R}}(\theta, \phi)$\\[4pt]
                         \hline
                         $\begin{aligned}
                        &\scalebox{0.001}{~}\\&\frac{1}{2}\\[7pt] -&\frac{1}{2} \\[7pt] &\frac{1}{2}\\[7pt] -&\frac{1}{2}
                        \end{aligned}$
                        &
                         $\begin{aligned}
                        &\scalebox{0.001}{~}\\&\frac{1}{2}\\[7pt] &\frac{1}{2} \\[7pt] -&\frac{1}{2}\\[7pt] -&\frac{1}{2}
                        \end{aligned}$
                        &
                         $\begin{aligned}
                        &\scalebox{0.001}{~}\\
                        \pm&\sum_j\frac{2j+1}{\sqrt{2}} \mathcal{D}^j(R) ~f^V_{-3}(R(I,j=l\pm\frac{1}{2}))~  d^j_{\frac{3}{2} \frac{1}{2}}(\theta) e^{-2i\phi}\\
                        &\sum_j\frac{2j+1}{\sqrt{2}} \mathcal{D}^j(R)~ f^V_{-3}(R(I,j=l\pm\frac{1}{2}))~  d^j_{\frac{3}{2} -\frac{1}{2}}(\theta)e^{-i\phi}\\
                        \mp&\sum_j\frac{2j+1}{\sqrt{2}} \mathcal{D}^j(R)~ f^V_{-1}(R(I,j=l\pm\frac{1}{2}))~  d^j_{\frac{1}{2} \frac{1}{2}}(\theta)e^{-i\phi}\\
                        -&\sum_j\frac{2j+1}{\sqrt{2}} \mathcal{D}^j(R)~ f^V_{-1}(R(I,j=l\pm\frac{1}{2}))~  d^j_{\frac{1}{2} -\frac{1}{2}}(\theta)\\
                        \end{aligned}$
                        &
                         $\begin{aligned}
                         &\scalebox{0.01}{~}\\
                        -&\sum_j\frac{2j+1}{\sqrt{2}} \mathcal{D}^j(R)~ f^V_{-1}(R(I,j=l\pm\frac{1}{2}))~  d^j_{\frac{1}{2} -\frac{1}{2}}(\theta)\\
                        \pm&\sum_j\frac{2j+1}{\sqrt{2}} \mathcal{D}^j(R)~ f^V_{-1}(R(I,j=l\pm\frac{1}{2}))~  d^j_{\frac{1}{2} \frac{1}{2}}(\theta)e^{i\phi}\\
                        -&\sum_j\frac{2j+1}{\sqrt{2}} \mathcal{D}^j(R)~ f^V_{-3}(R(I,j=l\pm\frac{1}{2}))~  d^j_{\frac{3}{2} -\frac{1}{2}}(\theta)e^{i\phi}\\
                        \pm&\sum_j\frac{2j+1}{\sqrt{2}} \mathcal{D}^j(R) ~f^V_{-3}(R(I,j=l\pm\frac{1}{2}))~  d^j_{\frac{3}{2} \frac{1}{2}}(\theta) e^{2i\phi}
                        \end{aligned}$
                        \\\hline
                        &  &$\tilde{F}_{\lambda_2 \lambda_1}^{e_{-}}(\theta, \phi)$& $\tilde{F}_{\lambda_2 \lambda_1}^{e_{+}}(\theta, \phi)$\\[4pt]
                         \hline
                         $\begin{aligned}
                        &\scalebox{0.001}{~}\\&\frac{1}{2}\\[7pt] -&\frac{1}{2} \\[7pt] &\frac{1}{2}\\[7pt] -&\frac{1}{2}
                        \end{aligned}$
                        &
                         $\begin{aligned}
                        &\scalebox{0.001}{~}\\&\frac{1}{2}\\[7pt] &\frac{1}{2} \\[7pt] -&\frac{1}{2}\\[7pt] -&\frac{1}{2}
                        \end{aligned}$
                        &
                         $\begin{aligned}
                        &\scalebox{0.05}{~}\\
                        &\frac{|\mathbf k|}{\sqrt{-k^2}}\sum_j\frac{2j+1}{\sqrt{2}} \mathcal{D}^j(R)~ f^{V(-)}_{0+}(R(I,j=l\pm\frac{1}{2}))~  d^j_{\frac{1}{2} \frac{1}{2}}(\theta)e^{-i\phi}\\
                        \pm&\frac{|\mathbf k|}{\sqrt{-k^2}}\sum_j\frac{2j+1}{\sqrt{2}} \mathcal{D}^j(R)~ f^{V(-)}_{0+}(R(I,j=l\pm\frac{1}{2}))~  d^j_{\frac{1}{2} -\frac{1}{2}}(\theta)\\
                        \pm&\frac{|\mathbf k|}{\sqrt{-k^2}}\sum_j\frac{2j+1}{\sqrt{2}} \mathcal{D}^j(R)~ f^{V(-)}_{0+}(R(I,j=l\pm\frac{1}{2}))~  d^j_{\frac{1}{2} -\frac{1}{2}}(\theta)\\
                        -&\frac{|\mathbf k|}{\sqrt{-k^2}}\sum_j\frac{2j+1}{\sqrt{2}} \mathcal{D}^j(R)~ f^{V(-)}_{0+}(R(I,j=l\pm\frac{1}{2}))~  d^j_{\frac{1}{2} \frac{1}{2}}(\theta)e^{i\phi}\end{aligned}$
                        &
                         $\begin{aligned}
                         &\scalebox{0.05}{~}\\
                        &\frac{|\mathbf k|}{\sqrt{-k^2}}\sum_j\frac{2j+1}{\sqrt{2}} \mathcal{D}^j(R)~ f^{V(+)}_{0+}(R(I,j=l\pm\frac{1}{2}))~  d^j_{\frac{1}{2} \frac{1}{2}}(\theta)e^{-i\phi}\\
                        \pm&\frac{|\mathbf k|}{\sqrt{-k^2}}\sum_j\frac{2j+1}{\sqrt{2}} \mathcal{D}^j(R)~ f^{V(+)}_{0+}(R(I,j=l\pm\frac{1}{2}))~  d^j_{\frac{1}{2} -\frac{1}{2}}(\theta)\\
                        \pm&\frac{|\mathbf k|}{\sqrt{-k^2}}\sum_j\frac{2j+1}{\sqrt{2}} \mathcal{D}^j(R)~ f^{V(+)}_{0+}(R(I,j=l\pm\frac{1}{2}))~  d^j_{\frac{1}{2} -\frac{1}{2}}(\theta)\\
                        -&\frac{|\mathbf k|}{\sqrt{-k^2}}\sum_j\frac{2j+1}{\sqrt{2}} \mathcal{D}^j(R)~ f^{V(+)}_{0+}(R(I,j=l\pm\frac{1}{2}))~  d^j_{\frac{1}{2} \frac{1}{2}}(\theta)e^{i\phi}\end{aligned}$
                        \\
                        \end{tabular}
                        \end{ruledtabular}
                        \end{table*}
            \begin{table*}
        \centering
        \caption{Axial helicity amplitudes of resonant interaction. }
        \label{res_HAA}
        \renewcommand{\arraystretch}{1.3}
        \begin{ruledtabular}
         \begin{tabular}{c|c|c|c}
          $\lambda_2$& $\lambda_1$ &$\tilde{G}_{\lambda_2 \lambda_1}^{e_{L}}(\theta, \phi)$& $\tilde{G}_{\lambda_2 \lambda_1}^{e_{R}}(\theta, \phi)$\\[4pt]
         \hline
         $\begin{aligned}
        &\scalebox{0.001}{~}\\&\frac{1}{2}\\[7pt] -&\frac{1}{2} \\[7pt] &\frac{1}{2}\\[7pt] -&\frac{1}{2}
        \end{aligned}$
        &
         $\begin{aligned}
        &\scalebox{0.001}{~}\\&\frac{1}{2}\\[7pt] &\frac{1}{2} \\[7pt] -&\frac{1}{2}\\[7pt] -&\frac{1}{2}
        \end{aligned}$
        &
         $\begin{aligned}
        &\scalebox{0.001}{~}\\
        \mp&\sum_j\frac{2j+1}{\sqrt{2}} \mathcal{D}^j(R) ~f^A_{-3}(R(I,j=l\pm\frac{1}{2}))~  d^j_{\frac{3}{2} \frac{1}{2}}(\theta) e^{-2i\phi}\\
        -&\sum_j\frac{2j+1}{\sqrt{2}} \mathcal{D}^j(R)~ f^A_{-3}(R(I,j=l\pm\frac{1}{2}))~  d^j_{\frac{3}{2} -\frac{1}{2}}(\theta)e^{-i\phi}\\
        \pm&\sum_j\frac{2j+1}{\sqrt{2}} \mathcal{D}^j(R)~ f^A_{-1}(R(I,j=l\pm\frac{1}{2}))~  d^j_{\frac{1}{2} \frac{1}{2}}(\theta)e^{-i\phi}\\
        &\sum_j\frac{2j+1}{\sqrt{2}} \mathcal{D}^j(R)~ f^A_{-1}(R(I,j=l\pm\frac{1}{2}))~  d^j_{\frac{1}{2} -\frac{1}{2}}(\theta)\\
        \end{aligned}$
        &
         $\begin{aligned}
         &\scalebox{0.01}{~}\\
        -&\sum_j\frac{2j+1}{\sqrt{2}} \mathcal{D}^j(R)~ f^A_{-1}(R(I,j=l\pm\frac{1}{2}))~  d^j_{\frac{1}{2} -\frac{1}{2}}(\theta)\\
        \pm&\sum_j\frac{2j+1}{\sqrt{2}} \mathcal{D}^j(R)~ f^A_{-1}(R(I,j=l\pm\frac{1}{2}))~  d^j_{\frac{1}{2} \frac{1}{2}}(\theta)e^{i\phi}\\
        -&\sum_j\frac{2j+1}{\sqrt{2}} \mathcal{D}^j(R)~ f^A_{-3}(R(I,j=l\pm\frac{1}{2}))~  d^j_{\frac{3}{2} -\frac{1}{2}}(\theta)e^{i\phi}\\
        \pm&\sum_j\frac{2j+1}{\sqrt{2}} \mathcal{D}^j(R) ~f^A_{-3}(R(I,j=l\pm\frac{1}{2}))~  d^j_{\frac{3}{2} \frac{1}{2}}(\theta) e^{2i\phi}
        \end{aligned}$
        \\\hline
        &  &$\tilde{G}_{\lambda_2 \lambda_1}^{e_{-}}(\theta, \phi)$& $\tilde{G}_{\lambda_2 \lambda_1}^{e_{+}}(\theta, \phi)$\\[4pt]
         \hline
         $\begin{aligned}
        &\scalebox{0.001}{~}\\&\frac{1}{2}\\[7pt] -&\frac{1}{2} \\[7pt] &\frac{1}{2}\\[7pt] -&\frac{1}{2}
        \end{aligned}$
        &
         $\begin{aligned}
        &\scalebox{0.001}{~}\\&\frac{1}{2}\\[7pt] &\frac{1}{2} \\[7pt] -&\frac{1}{2}\\[7pt] -&\frac{1}{2}
        \end{aligned}$
        &
         $\begin{aligned}
        &\scalebox{0.05}{~}\\
        -&\frac{|\mathbf k|}{\sqrt{-k^2}}\sum_j\frac{2j+1}{\sqrt{2}} \mathcal{D}^j(R)~ f^{A(-)}_{0+}(R(I,j=l\pm\frac{1}{2}))~  d^j_{\frac{1}{2} \frac{1}{2}}(\theta)e^{-i\phi}\\
         \mp&\frac{|\mathbf k|}{\sqrt{-k^2}}\sum_j\frac{2j+1}{\sqrt{2}} \mathcal{D}^j(R)~ f^{A(-)}_{0+}(R(I,j=l\pm\frac{1}{2}))~  d^j_{\frac{1}{2} -\frac{1}{2}}(\theta)\\
         \pm&\frac{|\mathbf k|}{\sqrt{-k^2}}\sum_j\frac{2j+1}{\sqrt{2}} \mathcal{D}^j(R)~ f^{A(-)}_{0+}(R(I,j=l\pm\frac{1}{2}))~  d^j_{\frac{1}{2} -\frac{1}{2}}(\theta)\\
        -&\frac{|\mathbf k|}{\sqrt{-k^2}}\sum_j\frac{2j+1}{\sqrt{2}} \mathcal{D}^j(R)~ f^{A(-)}_{0+}(R(I,j=l\pm\frac{1}{2}))~  d^j_{\frac{1}{2} \frac{1}{2}}(\theta)e^{i\phi}\end{aligned}$
        &
         $\begin{aligned}
         &\scalebox{0.05}{~}\\
        -&\frac{|\mathbf k|}{\sqrt{-k^2}}\sum_j\frac{2j+1}{\sqrt{2}} \mathcal{D}^j(R)~ f^{A(+)}_{0+}(R(I,j=l\pm\frac{1}{2}))~  d^j_{\frac{1}{2} \frac{1}{2}}(\theta)e^{-i\phi}\\
         \mp&\frac{|\mathbf k|}{\sqrt{-k^2}}\sum_j\frac{2j+1}{\sqrt{2}} \mathcal{D}^j(R)~ f^{A(+)}_{0+}(R(I,j=l\pm\frac{1}{2}))~  d^j_{\frac{1}{2} -\frac{1}{2}}(\theta)\\
         \pm&\frac{|\mathbf k|}{\sqrt{-k^2}}\sum_j\frac{2j+1}{\sqrt{2}} \mathcal{D}^j(R)~ f^{A(+)}_{0+}(R(I,j=l\pm\frac{1}{2}))~  d^j_{\frac{1}{2} -\frac{1}{2}}(\theta)\\
        -&\frac{|\mathbf k|}{\sqrt{-k^2}}\sum_j\frac{2j+1}{\sqrt{2}} \mathcal{D}^j(R)~ f^{A(+)}_{0+}(R(I,j=l\pm\frac{1}{2}))~  d^j_{\frac{1}{2} \frac{1}{2}}(\theta)e^{i\phi}\end{aligned}$
        \\
        \end{tabular}
        \end{ruledtabular}
        \end{table*}

                       \nocite{*}
        \bibliographystyle{unsrt} 
     \bibliography{bib_ma28}               
                        \end{document}